\documentclass[11pt,a4paper]{article}
\usepackage{jheppub}
 
\hypersetup{colorlinks=true,backref=true,linkcolor=black,anchorcolor=black,citecolor=black,filecolor   =black,menucolor=black,pagecolor=black,urlcolor=black}
 
\usepackage{amsfonts}
\usepackage{amsmath}
\usepackage{amssymb}
\usepackage{latexsym}
\usepackage{graphicx}

\makeatletter
\def\@fpheader{}
\makeatother

\numberwithin{equation}{section}

\def\nn{\nonumber}

\newcommand{\ord}[1]{{\scriptscriptstyle (#1)}}
\def\a{\alpha}
\def\b{\beta}
\def\c{\gamma}
\def\d{\delta}

\def\a{\alpha}
\def\b{\beta}
\def\c{\gamma}
\def\d{\delta}
\def\e{\epsilon}\def\ve{\varepsilon}
\def\F{\Phi}

\def\s{\sigma}

\def\o{\omega}\def\O{\Omega}

\def\gh{\mathfrak{h}}
\def\gf{\mathfrak{f}}
\def\gk{\mathfrak{k}}

\def\HSSA{{\underline{A}}}
\def\HSSB{{\underline{B}}}
\def\HSSC{{\underline{C}}}

\def\IZ{{\mathbb Z}}
\def\IC{{\mathbb C}}

\def\IF{{\mathbb F}}

\def\um{{\underline{\mu}}}
\def\umd{{\underline{\dot\mu}}}

\def\u|#1|#2{u_{#1}{}^{#2}}
\def\ur|#1|#2{u_{#1}{}^{\underline{#2}}}
\def\lu|#1|#2{u_{\underline{#1}}{}^{#2}}
\def\ua{{\underline{\alpha}}}

\def\uad{{\underline{\dot\alpha}}}

\def\cG{{\cal G}}

\def\ad{{\dot\alpha}}
\def\bd{{\dot\beta}}
\def\cd{{\dot\gamma}}
\def\dd{{\dot\delta}}
\def\ed{{\dot\epsilon}}

\def\ha{{\hat{\alpha}}}
\def\hb{{\hat{\beta}}}

\def\cE{\mathcal{E}}
\def\cN{\mathcal{N}}
\def\cV{\mathcal{V}}

\def\csN{{\fontsize{9.35pt}{9pt}\selectfont \mbox{$\cN$} \fontsize{12.35pt}{12pt}\selectfont }}
\def\cssN{{\fontsize{6.35pt}{6pt}\selectfont \mbox{$\cN$} \fontsize{12.35pt}{12pt}\selectfont }}
\def\csssN{{\fontsize{4.35pt}{4pt}\selectfont \mbox{$\cN$} \fontsize{12.35pt}{12pt}\selectfont }}

\def\csI{{\fontsize{9.35pt}{9pt}\selectfont \mbox{$\cal I$} \fontsize{12.35pt}{12pt}\selectfont }}
\def\csJ{{\fontsize{9.35pt}{9pt}\selectfont \mbox{$\cal J$} \fontsize{12.35pt}{12pt}\selectfont }}

\def\gI{{\fontsize{6.35pt}{6pt}\selectfont \mbox{$\cal I$} \fontsize{12.35pt}{12pt}\selectfont }}
\def\gJ{{\fontsize{6.35pt}{6pt}\selectfont \mbox{$\cal J$} \fontsize{12.35pt}{12pt}\selectfont }}
\def\gK{{\fontsize{6.35pt}{6pt}\selectfont \mbox{$\cal K$} \fontsize{12.35pt}{12pt}\selectfont }}
\def\gL{{\fontsize{6.35pt}{6pt}\selectfont \mbox{$\cal L$} \fontsize{12.35pt}{12pt}\selectfont }}
\def\gP{{\fontsize{6.35pt}{6pt}\selectfont \mbox{$\cal P$} \fontsize{12.35pt}{12pt}\selectfont }}
\def\gQ{{\fontsize{6.35pt}{6pt}\selectfont \mbox{$\cal Q$} \fontsize{12.35pt}{12pt}\selectfont }}
\def\gM{{\fontsize{6.35pt}{6pt}\selectfont \mbox{$\cal M$} \fontsize{12.35pt}{12pt}\selectfont }}
\def\gN{{\fontsize{6.35pt}{6pt}\selectfont \mbox{$\cal N$} \fontsize{12.35pt}{12pt}\selectfont }}

\newcommand{\eprint}[1]{{\href{http://arxiv.org/abs/#1}{[arXiv:\,\texttt{#1}}]}}
\newcommand{\eprintN}[1]{{\href{http://arxiv.org/abs/#1}{[arXiv:\,\texttt{#1\,[hep-th]}]}}}

\def\ie{{\it i.e.}\ }
\def\eg{{\it e.g.}\ }
\def\cf{{\it cf.}\ }

\def\be{\begin{equation}}
\def\ee{\end{equation}}
\def\bea{\begin{eqnarray}}
\def\eea{\end{eqnarray}}
\newcommand{\CR}{\nonumber \\*}

\title{The vanishing volume of \boldmath $D=4$ superspace}

\author[a]{Guillaume Bossard,}
\author[b]{P.S. Howe,}
\author[c,d,e]{K.S. Stelle,}
\author[d,f,g]{Pierre Vanhove}
\affiliation[a]{Centre de Physique Th{\'e}orique, Ecole Polytechnique, CNRS\\ 91128 Palaiseau Cedex, France}
\affiliation[b]{Department of Mathematics, King's College, University of London\\ Strand, London WC2R 2LS, UK}
\affiliation[c]{Theoretical Physics Group, Imperial College London\\ \quad Prince Consort Road, London SW7 2AZ, UK}
\affiliation[d]{Kavli Institute for Theoretical Physics, University of California\\ Santa Barbara CA 93106, USA}
\affiliation[e]{TEO/CBPF, Rua Dr.\ Xavier Sigaud 150\\
cep 22290-180, Rio de Janeiro (RJ), Brazil}
\affiliation[f]{Institut des Hautes Etudes Scientifiques\\
Le Bois-Marie, F-91440 Bures-sur-Yvette, France}
\affiliation[g]{Institut de Physique Th{\'e}orique, CEA, IPhT, F-91191 Gif-sur-Yvette, France\\
CNRS, URA 2306, F-91191 Gif-sur-Yvette, France}

\emailAdd{bossard@cpht.polytechnique.fr}
\emailAdd{paul.howe@kcl.ac.uk}
\emailAdd{k.stelle@imperial.ac.uk}
\emailAdd{pierre.vanhove@cea.fr}

\abstract{The volume of on-shell $D=4$, $\cN=8$ superspace is shown to vanish. Despite this, it is shown that there is a fully supersymmetric and
duality-invariant candidate $\nabla^8R^4$ counterterm corresponding to an anticipated seven-loop
logarithmic  divergence in $D=4$.  We construct this counterterm explicitly and also give the complete
nonlinear extension  of the $1/8$-BPS $\nabla^6R^4$ invariant. Similar results are derived for $\cN=4,5\ \&\ 6$.}

\begin{document}

\begin{flushright}   \small    CBPF-NF-011/11   \\   CPHT-RR045.0511\\
  IHES/P/11/14\\ Imperial/TP/11/KSS/02\\ IPHT-T-11/124\\ KCL-MTH-11-11 \\NSF-KITP-11-069 
\end{flushright}
\vspace{0.5cm}

\maketitle

\section{Introduction}
\label{sec:intro}

The problem of ultraviolet divergences in supergravity has attracted the attention of theorists since the origins of the theory. Early on, it was realised that candidate supersymmetric counterterms, non-vanishing subject to the classical equations of motion, exist starting from the 3-loop level in $D=4$, where they would generically be of $(\textrm{curvature})^4$ structure \cite{Deser:1977nt,Deser:1978br,Kallosh:1980fi,Howe:1981xy}. 
It was noted at the time that, with respect to the full ``on-shell'' supersymmetry of the $\cN>4$ extended supergravities, these $(\textrm{curvature})^4$ counterterms need to be expressed as subsurface integrals over the full superspace, \ie as ``BPS'', or ``F-term'' invariants. But according to the understanding at the time of the possible linearly realisable ``off-shell'' supersymmetry, which is less than the full on-shell degree, they appeared to be expressible as full superspace integrals of the linearly realisable off-shell supersymmetry and thus were not thought to violate applicable non-renormalisation theorems \cite{Howe:1983sr}. Thus, despite their BPS subsurface-integral structure, the ultraviolet divergences of $D=4$ supergravity looked set to begin at the 3-loop level, provided the maximal off-shell linearly realisable supersymmetry corresponded to just half the full on-shell degree. Should the linearly realisable supersymmetry turn out to be more than half the full on-shell degree, \eg through a harmonic superspace formulation, the divergence onset loop order would correspondingly rise \cite{Howe:2002ui,Bossard:2009sy}.

In case the non-renormalisation theorems for BPS invariants turned out to be stronger than anticipated, it was also noted in the 1980s that full-superspace integral invariants would in any case be available starting at the 7-loop or 8-loop order \cite{Howe:1980th,Kallosh:1980fi}. The constraints of continuous duality symmetries such as $E_{7(7)}$ for the maximal $\cN=8$ theory were recognised to be important as well. The 8-loop full-superspace counterterm was recognised to be manifestly duality invariant. But it was also anticipated that a duality-invariant counterterm could exist already at 7 loops, where na{\"\i}ve power counting in $D=4$ gives an expectation of a dimension 16 counterterm, corresponding  to the $\int d^{32}\theta$ full-superspace integral for maximal supergravity. An obvious candidate for such a dimension-16 duality invariant non-BPS counterterm was the full volume of the $\cN=8$ superspace, $\int d^4x d^{32}\theta E$, where $E$ is the Berezinian determinant of the supervielbein.

For the $\cN\le3$ lesser extended supergravities, it was recognised that the volume of superspace vanishes subject to the classical supergravity  field equations, for a series of specific reasons. For $\cN=1$, the volume of superspace gives the dimension-2 supergravity action \cite{Wess:1978bu}, and thus vanishes on-shell for the ``non-gauged'' Poincar{\'e} supergravities without a cosmological constant. Indeed, in the ``new minimal'' auxiliary-field formulation, the superspace volume  vanishes even off-shell \cite{Howe:1981et,Sohnius:1982fw}. In the $\cN=2$ case, the vanishing of the superspace volume was expected because the corresponding dimension-4 
$(\textrm{curvature})^2$ invariants are constrained by the Gauss-Bonnet identity to be equivalent, up to a total divergence, to quadratic expressions in the Ricci tensor or its trace, thus leading to on-shell vanishing counterterms. The vanishing of the $\cN=2$ superspace volume was confirmed explicitly in~\cite{SokatchevTD} by reducing it to a chiral integral. In the $\cN=3$ case, it was similarly known that there are no dimension-6 counterterms (corresponding to $(\textrm{curvature})^3$ structures) that are non-vanishing subject to the classical equations of motion \cite{Grisaru:1976nn}. Aside from these rather transparent low-$\csN$ supergravity cases, however, there seemed to be no particular reason why the superspace volume should vanish for the higher-$\csN$ extended supergravities.

In the meantime,  computational techniques have improved dramatically,
and  much more  is  now  known from  explicit  calculations about  the
ultraviolet divergences of supergravity (see \cite{Bern:2011qn} for
a recent review).  The result is that, despite the anticipation of first maximal supergravity divergences at 3 loops,  ultraviolet cancellations  turn  out to  continue unabated  in $D=4$ and also in $D=5$ at the 3-loop  \cite{Bern:2007hh,Bern:2008pv} and at the 4-loop levels \cite{Bern:2009kd}.\footnote{Owing to the on-shell
  conditions, no non-vanishing  4-loop divergences could have appeared
  in the  $\cN=8$, $D=4$ four-graviton  amplitude. Moreover, nonlinear
  $\nabla^2R^4$  and   $R^5$  invariants  were  ruled   out  in  Ref.\
  \cite{Drummond:2003ex}. A  discussion of the  kinematic structure of
  four-point  counterterms in $D=4$  non-maximal supergravity  will be
  given in Appendix~\ref{sec:Kinematics}.} This clearly required revisiting the analysis of the non-renormalisation theorems. Indeed, although the earlier 1980s non-renormalisation analysis had relied upon the known off-shell linearly realisable degree of supersymmetry, it turns out that the full on-shell supersymmetry imposes further constraints that were not initially recognised. Even though the full on-shell supersymmetry involves nonlinear transformations and is thus subject to complicated transformation renormalisations, the corresponding Ward identities, expressed using BRST algebraic renormalisation techniques, show that the $(\textrm{curvature})^4$ counterterm previously anticipated at the $D=4$ 3-loop level, is actually ruled out \cite{Bossard:2009sy}. Similarly, the BPS counterterms expected at loop orders up to 6 were brought under suspicion.

Another aspect of the BPS counterterms that was missed in the original 1980s analysis is their delicacy with respect to the continuous duality symmetries. Originally, the only analysis that could be carried out used linearised $\cN=8$ supersymmetry transformations, focusing purely on the leading 4-particle level of the candidate counterterms. At this leading order, the $(\textrm{curvature})^4$ candidate passed the only available test of duality invariance, namely invariance  under constant shifts of the 70 scalar fields. This happened because at the 4-point level all scalar fields in the invariant are covered by derivatives \cite{Howe:1983sr}. But little was known at the time about the full nonlinear structure of the $(\textrm{curvature})^4$ candidate. This became much clearer recently, however, through relations between counterterms obtained via dimensional reduction, starting from field-theory limits of string-theory amplitudes \cite{Elvang:2010kc,Green:2010sp,Green:2010kv} or purely within supergravity \cite{Bossard:2010bd}. The result is that, contrary to the initial 1980s impression that the $(\textrm{curvature})^4$ counterterm might be $E_{7(7)}$ invariant, it in fact turns out to fail this test at the nonlinear level, owing to scalar-field ``dressings'' of the purely gravitational $(\textrm{curvature})^4$ term. There are just two other linearised BPS invariants in $D=4,\cN=8$ supergravity, at 5 and 6-loops \cite{Drummond:2003ex,Elvang:2010jv,Drummond:2010fp}, but these also turn out to be incompatible with $E_{7(7)}$ invariance at the nonlinear level~\cite{Beisert:2010jx,Bossard:2010bd}. So, none of the $D=4$ F-term invariants can correspond to divergences, because it is now known that $E_{7(7)}$ can be preserved in the quantum perturbation theory \cite{Bossard:2010dq}. These duality invariance requirements end up invalidating the previously-thought-acceptable BPS counterterms at 3 through 6 loops in $D=4$ maximal supergravity \cite{Bossard:2010bd,Beisert:2010jx,Green:2010sp,Green:2010kv}.

Consequently, the candidate counterterm at the 7-loop order assumes a greater importance than it was previously accorded: it is now the leading candidate for a $D=4$ maximal supergravity divergence. So the question of its structure becomes of key importance, and in particular the question whether it can in fact be written as the full-superspace volume of the maximal theory. This is the question that we will address in the present paper.

We will prove two main results: firstly, that the volume of superspace actually vanishes on-shell for any $\cN$, and secondly, despite this, that there are nevertheless duality-symmetric invariants of the same dimension, schematically of the form $\nabla^{2\cssN-8}R^4$. These invariants correspond to possible counterterms at the $(\csN-1)$-loop level.

We do not currently know any obvious {\it a priori} reason why the $D=4$ superspace volumes should vanish. The proof that it does relies on  harmonic superspace methods to reduce full superspace integrals to integrals over superspaces with four fewer odd coordinates. A quick way of understanding the result is to consider how one might integrate an unconstrained scalar superfield $\Phi$ over the reduced superspace using an appropriate projection operator. For example, in off-shell minimal $\cN=1$ supergravity the chiral projection operator is $\bar D^2+ S$, where $\bar D^2:=\ve^{\ad\bd} \bar D_\ad\bar D_\bd$ and $S$ is a chiral superfield whose leading component is the complex auxiliary scalar. So the integral of $\F$ over the full superspace is equal to the integral of $(\bar D^2+ S)\F$ over chiral superspace. If we take $\F=1$, this integral, which is just the volume of superspace, need not vanish. On the other hand, on-shell one has $S=0$, and so the volume vanishes on-shell. In the new-minimal formalism, however, a $U(1)$ connection is included in the covariant derivative, the superfield $S$ vanishes and the projector is simply $\bar D^2$ so that the volume vanishes even off-shell \cite{Howe:1981et,Sohnius:1982fw}.

In $\cN$-extended on-shell supergravity it turns out, as we shall see, that one can select one undotted and one dotted covariant spinor covariant derivative, say  $D_\a^1$ and $\bar D_{\ad \cssN}$,   that anticommute with each other when acting on scalar superfields. There are therefore G-analytic superfields \cite{Galperin:1980fg} that are, by definition, annihilated by these derivatives and that can be thought of as ``generalised chiral'' superfields. It turns out, as we shall prove later,  that $(D^1)^2$ commutes with  $(\bar D_{\cssN})^2$ and that the associated projection operators are $(D^1)^2(\bar D_{\cssN})^2$. This means that we can integrate a scalar superfield $\F$ over a superspace with four fewer odd coordinates and that, as a corollary, the volume of the full superspace must vanish because we can write it as a sub-superspace integral of $(D^1)^2(\bar D_{\cssN})^2$ acting on the constant superfield integrand $\Phi=1$. 

In  Section~\ref{sec:normal},   we  define   $(\csN,1,1)$  harmonic
superspace  for supergravity theories and show  that   the  volume of  superspace vanishes for all $\cN$. In
Section~\ref{sec:volume},  we  show   that  full  superspace  integrals
can be reduced to integrals with respect to  the $(\csN,1,1)$ harmonic superspace measure. In
Section~\ref{sec:invariants}, using this harmonic measure, we construct
fully supersymmetric and duality-invariant $(\csN-1)$-loop counterterms of general structure $\nabla^{2(\cssN-4)}R^4$.
In addition, we construct nonlinear  versions  of the 
non-duality invariant $1/\csN$-BPS supersymmetry invariants of general structure
$\nabla^{2(\cssN-5)}R^4$ and clarify the classification of duality invariant $\cN$-loop candidate counterterms.  Section~\ref{sec:conclusion} contains our
conclusions. Conventions and details about extended on-shell superspace
are given  in Appendix~\ref{sec:superspace}. Appendix~\ref{sec:Kinematics}
contains an analysis of the kinematic structure of the derivative
expansion appearing in the four-graviton amplitudes in supergravity.

\section{Superspace formalism}
\label{sec:normal}
\subsection{Standard superspace}
In $D=4$, $\cN$-extended superspace, $M$, is a supermanifold with 4 even and $4\cN$ odd dimensions;  local coordinates are denoted by
$z^M=(x^m,\theta^{\um},\bar\theta^{\umd})$ where $x^m$ are the even, spacetime, coordinates and the thetas are the odd coordinates. The preferred basis forms are  $E^A:=(E^a,    E^{\ua}, E^{\uad})$ with $E^\ua=E^\a_i$, $E^\uad=E^{\dot\alpha i}$.   The index $i$ runs from 1 to $\cN$, $\a$ and $\ad$ are two-component spinor indices and underlined indices combine internal and spinor ones. The structure group, under which the preferred frames transform, is $SL(2,\IC)\times U(\csN)$, with the former factor acting on the vector index $a$ in the usual way. The connection, torsion and curvature are defined as usual with
\bea
T^A&=& DE^A:= dE^A + E^B\O_B{}^A \nonumber \ , \\
R_A{}^B &=& d\O_A{}^B + \O_A{}^C\O_C{}^B\ .
\eea
Because the structure group is purely even it follows that the mixed, even-odd, components of the connection one-forms, $\O_A{}^B$, and the curvature two-forms, $R_A{}^B$, are zero. The dimension-zero torsion does not involve the connection and takes the same form as it does in flat superspace, namely
\bea
T_{\a\b}^{i\,j\,c}&=&0 \nonumber\ , \\
T_{\a\bd j}^{i\ \ c}&=&-i\d^i_j(\s^c)_{\a\bd}\ .
\eea
These equations, together with the conventional constraints that allow one to choose the connection and the vectorial basis $E_a$ \cite{Gates:1979wg}, determine the conformal constraints that were discussed in~\cite{Howe:1981gz}, to which paper we refer for further details. (We also collect some useful results in Appendix~\ref{sec:superspace}.) The Bianchi identities corresponding to these constraints were solved in detail in~\cite{Howe:1981gz}; we note here that the dimension one-half torsion components are zero except for
\be
T_{\a\b}^{i\,j\cd k}=\ve_{\a\b}\bar\chi^{\cd ijk}
\ee
and its complex conjugate. The leading component of the field $\chi_{\a ijk}$ (in which roman-index sequences like $ijk$ are understood to be totally antisymmetric) denotes the 56 spin-one-half fields in the supergravity multiplet for $\cN=8$; there are additional independent spinors $\chi_\a^{ijklm}$ for $\cN=5,6$. The on-shell theory is completed at dimension one by specifying a number of superfields in terms of the physical component fields \cite{Brink:1979nt}. In addition to the geometrical fields, there are also spin-one field strengths and the scalars, the latter entering via a coset sigma model $K\backslash G$ which for $\cN=8$ is $(SU(8)/\IZ_2)\backslash E_{7(7)}$.

A key point about the above equations is that they are compatible with at most one $D_\a$ and one $\bar D_\ad$ being in involution; indeed, one could say that the dimension-zero torsion constraints are representation-preserving \cite{Gates:1979wg} for fields that are annihilated by such a set of derivatives. However, we clearly cannot pick out such a pair in ordinary superspace without breaking $U(\csN)$ symmetry, and for this reason we need to enlarge the setting to harmonic superspace.

\subsection{Harmonic superspace}\label{harmonicsspace}
Harmonic superspace (and the closely related projective superspace) is ordinary superspace augmented by an additional bosonic space that parametrises sets of mutually anticommuting fermionic derivatives \cite{Rosly:1982,Galperin:1984av,Karlhede:1984vr}. The spaces of most interest to us are the flag manifolds
\be \IF_{p,q}(\csN) \cong  \bigl( U(p)\times  U(\csN-q-p)\times U(q)\bigr)\backslash U(\csN) \ee
which parametrise the possible sets of $p$ undotted and $q$ dotted spinorial derivatives that anticommute on scalar fields \cite{Howe:1995md}.\footnote{There are more general internal flag spaces than those of (2.4) (see, e.g. \cite{Ivanov:1984ut} and \cite{Howe:1995md}), but they do not have the same geometric interpretation. Moreover, the associated harmonic analyticities are not compatible with the supergravity constraints.}
In our case, we need $p=q=1$ which gives $\IF_{1,1}(\csN)$. One way of working on such a coset space is to consider functions on the group $K\cong U(\csN)$ that are equivariant with respect to the isotropy group $H\cong U(p)\times  U(\csN-q-p)\times U(q)$, as advocated in the work of~\cite{Galperin:1984av,Galperin:2001uw}. In supergravity,\footnote{Other aspects of $D=4, \cN=2$ supergravity, including off-shell Poincar{\'e} supergravity, have been studied in harmonic superspace, \cf for example  \cite{Galperin:1987ek,Galperin:1987em}, and more recently in projective superspace \cite{Kuzenko:2008ep,Kuzenko:2009zu}. It has so far proved difficult to extend the off-shell Poincar{\'e} formalism to $\cN>2$ \cite{Sokzup}.} $U(\csN)$ is a gauge group so this means that in the equivariant formalism we should work on the principal $U(\csN)$ bundle which we will call $P$. We denote an element of $U(\csN)$ by $u^I{}_i$ where the local gauge group acts to the right and the isotropy group acts to the left. The inverse is denoted $u^i{}_I$. We can split the $I$ index according to the structure of the isotropy group: $I=(1,r,\csN)$, and we use $u$ or its inverse to convert $K$ indices to $H$ ones. In particular, for the fermionic derivatives, we have
\bea
D_\a^I&=&u^I{}_i D_\a^i =(D_\a^1, D_\a^r,D_\a^\cssN)\nonumber\\
\bar D_{\ad I}&=&u^i{}_I  \bar D_{\ad i} =(\bar D_{\ad 1},\bar D_{\ad r},\bar D_{\ad \cssN})\ .
\eea
One can immediately see that $D_\a^1$ and $\bar D_{\ad \cssN}$ anticommute among themselves, at least as far as the torsion is concerned, owing to the antisymmetry of $\chi_{\a ijk}$. (The curvature terms will be discussed shortly). In addition to the superspace derivatives, we also have the group derivatives $D^I{}_J$ which are simply the right-invariant vector fields on $K$; they obey the Lie algebra commutation relations for $U(\csN)$ and act in a simple fashion on $u$,
\be 
D^I{}_J u^K{}_k=\d^K_J u^I{}_k\ .\label{Durel}
\ee
(There is also a trace term for $SU(8)$.) These derivatives split into those corresponding to the isotropy algebra $\gh$, 
$(D^1{}_1,D^r{}_s,D^\cssN{}_\cssN)$ and the remainder corresponding to the coset directions $\gf$, where the Lie algebra $\gk$ of $K$ splits into $\gk\cong\gh\oplus \gf$. Since the coset space is complex, the latter divide into two complex conjugate sets: $(D^1{}_r,D^r{}_\cssN,D^1{}_\cssN)$ and $(D^r{}_1,D^\cssN{}_r,D^\cssN{}_1)$. 

In the principal bundle $P$ there is a Lie-algebra-valued one-form $\o$ that combines the Maurer--Cartan form on the group with the $U(\csN)$ connection on the base:
\be \o=du \, u^{-1} + u\O u^{-1}\ .\ee
A complete set of basis forms is then given by adding to these the basis vielbein forms on the supermanifold $M$. The dual basis vector fields are the right-invariant vectors fields on $K$ together with the horizontal lifts of the basis vectors on $M$, $\tilde E_A$. The latter are given by
\be \tilde E_A=E_A- \O_{A\ \,J}^{\ \ I} D^J{}_I\ .\ee
The    set   of   vector    fields   $(D_\a^1,\bar    D_{\ad   \cssN},
D^r{}_1,D^\cssN{}_1,D^\cssN{}_r)$ span a CR structure in the principal
bundle $P$, \ie an involutive, complex distribution that  has a
null intersection with the complex conjugate set. The proof of this is
given in~\cite{Hartwell:1994rp}; it depends on the details of the curvature tensor. The number of odd vector fields in this set cannot be increased for $\cN=5,6,8$, although one can have $(2,1)$ structures in $\cN=3,4$, and a $(2,2)$ structure in $\cN=4$.

Instead of working on $P$ it will turn out to be useful for the normal coordinate discussion to work directly on harmonic superspace $M_H$. This is the associated fibre bundle with fibre the coset space $F\cong H\backslash K$, where $F$ is the flag manifold, \ie $\IF_{1,1}(\csN)$, described above. To derive a convenient basis of forms on this space, one simply needs to split $\o$ into its isotropy and coset components, $\o=\o_\gh+\o_\gf$. The latter will be interpreted as a vertical vielbein while the former is a connection for $H$. The form basis is completed by the vielbein forms  from the base, but we have to contract the fermionic ones with $u$ or $u^{-1}$ so that they are not acted on by $K$ directly. Thus, $E^\a_I= E^\a_i u^i{}_I$ while $E^{\ad I}=u^I{}_i E^{\ad i}$. The resulting space has the structure group $SL(2,\IC)\times H$, although one should note that there has not been a choice of $U(\csN)$ gauge. One can work out the components of the torsion from the equation
\be d\o+\o^2= uRu^{-1}\ ,\ee
where $R$ is the $\mathfrak{k}\cong\mathfrak{u}(\csN)$ component of the curvature, simply by decomposing it into its isotropy and coset components. We have
\bea
D\o_\gf&=&-(\o_\gf\wedge\o_\gf)_\gf+(uRu^{-1})_\gf\nonumber\\
d\o_\gh+\o_\gh^2&=&-(\o_\gf\wedge\o_\gf)_\gh +(uRu^{-1})_\gh\ ,
\eea
where $D$ here denotes the exterior derivative which is covariant with respect to $H$. In these equations, we have fixed the gauge with respect to the isotropy group acting on $K$ so that $u$ should here be considered as a function of local coordinates, $t$ say, on $F$.  It will be useful to introduce a quantity $h(I)$ such that 
\be h(1) = 1\ , \quad h(r) = 0 \ , \quad h(\csN) = -1 \ . \ee
The coset indices are then pairs $I,J$ such that $h(I)\neq h(J)$ while the $H$-indices are pairs $I,J$ with $h(I)=h(J)$. The vielbein $V^I{}_J$ on $F$ is $(du \, u^{-1})_\gf$, and the corresponding quantity on $M_H$ is $\tilde V^I{}_J$ which is given by $\o_\gf=(du \, u^{-1}+u\O u^{-1})_\gf$. Thus,
\bea\label{e:VIJ}
V^I{}_J&=&d u^I{}_i \; u^i{}_J \nonumber\\
\tilde V^I{}_J&=&d u^I{}_i \; u^i{}_J +  u^I{}_i \O^i{}_j u^j{}_J\ ,
\eea
where in both of these expressions $h(I)\neq h(J)$. 
The full set of basis forms is thus $\tilde E^\HSSA=(\tilde V^I{}_J, E^a, E^\a_I,E^{\ad I})$. The torsion 2-form on $M_H$, $\tilde T^\HSSA$\,, is given by
\be \tilde T^a=T^a \ , \quad \tilde T^\alpha_I = T_i^\alpha  u^i{}_I+  E^\alpha_J\wedge \tilde V^J{}_I \ , \quad \tilde T^{I}{}_J = -\tilde V^I{}_K\wedge\tilde V^K{}_J +u^I{}_i R^i{}_j u^j{}_J\ , \ee
where $h(I)\neq h(J)\neq h(K)$\,.

We denote the vector fields on $F$ dual to $V^I{}_J$ by $d^I{}_J$; they are only defined for $h(I)\neq h(J)$. 
The complete set of vector fields dual to the basis forms consists of the $d^I{}_J$ together with the horizontal lifts of the basis vector fields of $M$ which we shall call $\tilde E_A$ with the understanding that the internal indices are capitalised. The full set is denoted $\tilde E_{\HSSA}=(\tilde E_A, d^I{}_J)$. One has
\be
\tilde E_A=E_A-\O_{A,\ \,J}^{\ \ I} d^J{}_I\ .
\ee
The combination $\O^I{}_J d^J{}_I$ (where the sum runs only over indices for which $h(I)\neq h(J)$) can be rewritten as $\O^i{}_j K^j{}_i$, where the $K^i{}_j$ are the Killing vector fields on $F$ that generate the right action of $K$ on the coset. The graded commutator of two basis vector fields is
\be[\tilde E_\HSSA,\tilde E_\HSSB]=C_{\HSSA\,\HSSB}{}^\HSSC \tilde E_\HSSC:= 
\left(\tilde \O_{\HSSA,\HSSB}{}^\HSSC -(-1)^{\HSSA\HSSB} \tilde \O_{\HSSB,\HSSA}{}^\HSSC - \tilde T_{\HSSA\,\HSSB}{}^\HSSC\right)\tilde E_\HSSC\ .
\ee
In particular, for two fermionic indices, for example undotted ones, we have
\be
\{\tilde E_\a^I,\tilde E_\b^J\}=T_{\a\b}^{IJ\cd K} \tilde E_{\cd K} - R_{\a\b,\  \,L}^{I\,J\,K} d^L{}_K +{\rm connection\  terms}\ .
\ee  
The term involving the curvature here is a torsion term from the point of view of harmonic superspace. Note that the connection terms refer to $SL(2,\IC)\times H$ and so do not mix the indices $(1,r,\cN)$. This formula, together with those for mixed and undotted spinor indices, allows one to show that the subset of vector fields
\be 
  \label{e:Ehat}
  \hat  E_{\hat{A}}:=\{\tilde
E_{\alpha}^1,\tilde  E_{\dot\alpha\,\cssN},d^1{}_r,d^r{}_\cssN,d^1{}_\cssN\}\ , \qquad
2\leq r\leq \csN-1
\ee 
is in involution,
\be 
  \label{e:Eclose}
  \{\hat E_{\hat{A}}, \hat E_{\hat{B}}\}= C_{{\hat{A}}{\hat{B}}}{}^{\hat{C}} \, \hat E_{\hat{C}}\ , 
\ee 
and  is preserved under the action of the structure group $SL(2,\IC) \times U(1) \times U(\csN-2) \times U(1)$. The vector fields $(d^1{}_r,d^r{}_\cssN,d^1{}_\cssN )$ indeed close under commutation (they obey the commutation relations of a Heisenberg algebra) and can be thought of as being in essence the components of the anti-holomorphic Dolbeault exterior derivative $\bar\partial$ on the coset.  It is obvious that they commute with $\tilde E_{\alpha}^1$ and $\tilde  E_{\dot\alpha\,\cssN}$ because the relation~\eqref{Durel} is also valid for the $d^I{}_J$:
\be
d^I{}_J u^K{}_k=\d^K_J u^I{}_k\qquad{\rm (where}\  h(I)\neq h(J))\ .\label{Durel2}
\ee
It is also clear that the  torsion term vanishes for the commutator of
any  two  of  these  odd  basis  vector  fields  owing  to  the  total
antisymmetry of $\chi_{\a ijk}$ in the $ijk$ indices.

The curvature term also has the desired properties, as one can see from~\cite{Howe:1981gz}. Setting   $(\tilde E_\a^i,\tilde E_{\ad \cssN}):=\tilde E_\ha$\,, we need to show that~\cite{Hartwell:1994rp}
\be
R_{\ha\hb,\  \,\cssN}^{\ \ \ \ 1}=R_{\ha\hb,\  \,r}^{\ \ \ \ 1}=R_{\ha\hb,\  \,\cssN}^{\ \ \ \ r}=0\ ,
\ee
because these components of the curvature tensor couple to the derivatives $(d^r{}_1,d^\cssN{}_r,d^\cssN{}_1 )$ in the commutator $\{\tilde E_\ha,\tilde E_\hb\}$. It follows that this is indeed the case because
\be
R^{i\, j\ k}_{\a\b,\ \, l}=\d^i_l N_{\a\b}^{jk} + \d^j_l N_{\a\b}^{ik}\ ,
\label{2.20}
\ee
while 
\bea
R_{\a \bd j,\ \,l}^{i\ \ \ \, k}&=&-J^{ik}_{\a\bd, jl} + \d^i_j H^k_{\a\bd l} +\frac{1}{2}\d^k_l H^i_{\a\bd j} -\d^k_j H^i_{\a\bd l}-\d^i_lH^k_{\a\bd j} \nonumber\\
&&\phantom{=}- \Bigl( \frac{1}{2} \delta^i_j \delta_l^k + \delta_j^k \delta^i_l + \delta_l^i \delta_j^k \Bigr)  G_{\a\bd } \ ,
\eea
where the tensors $G$, $H$, $J$ and $N$ are given in Appendix \ref{sec:superspace}. (The curvature with two dotted spinor indices is the complex conjugate of the one with two undotted indices.) They are all bilinears in the fermion fields $\chi,\ \bar\chi$. 
To see more explicitly that the curvature has the desired properties, consider first  the undotted vector fields  $\tilde E_\a^1,\tilde E_\b^1$. In order for these  to be part of the involutive set~(\ref{e:Ehat}), we require 
\be \label{e:Rzero1} R^{1\, 1\ 1}_{\a\b,\ \, \cssN}=R^{1\,
    1\   1}_{\a\b,\   \,   r}=R^{1\,   1\  r}_{\a\b,\   \,   \cssN}=0\
  .\ee 
It is obvious that these conditions are satisfied owing to the presence of the Kronecker deltas in~(\ref{2.20}). A similar discussion is valid for the case of two dotted indices by complex conjugation. For the mixed index case, we need to show that
\be \label{e:Rzero2}  R_{\a  \bd  \cssN,\ \cssN}^{1\  \  \
    \,1}=R_{\a  \bd  \cssN,\  \,r}^{1\  \  \  \,1}=R_{\a  \bd  \cssN,\
    \,\cssN}^{1\ \ \ \, r}=0\ . \ee 
The tensor $J$ is not a problem because it is antisymmetric on both its upper and lower indices, while the terms involving $G$ and $H$ cannot be non-zero, again  because of the Kronecker deltas.

The explicit form of the involution equations (\ref{e:Eclose}) is therefore
\begin{eqnarray}
  \label{e:Ecom}
\nn \{\tilde E_\a^1,\tilde E_\b^1\}&=&
2\Omega_{(\a \b)}^{1\ \ \ \c} \,\tilde E_\c^1 + 2 \Omega_{(\a}^1{}^{1}{}_1 \,\tilde E_{\b)}^1 
-2\, N^{1r}_{\alpha\beta}\, d^1{}_r\\
\nn \{\tilde E_{\ad \cssN},\tilde E_{\bd \cssN} \}&=& -  2\Omega_{(\ad \cssN \, 
    \bd)}{}^{\cd}  \tilde    E_{\cd    \cssN}   -2    \Omega_{(\ad
    \cssN}{}^\cssN{}_\cssN    \,     \tilde    E_{\bd )   \cssN}    -2
  \, \bar N_{\ad\bd\,\cssN\, r}\, d^r{}_\cssN \\
    \{\tilde E_\a^1,\tilde E_{\bd   \cssN}\}&=& -\Omega_{\a}^1{}_{\, \bd}{}^{\cd}\,\tilde E_{\cd \cssN} 
 +  \Omega_{\bd   \cssN\,\a}{}^{\c}\,\tilde E_{\c}^1
- \Omega_{\a}^1{}^{\cssN}{}_{\cssN}\,\tilde
 E_{\bd \cssN}  +   \Omega_{\bd   \cssN}{}^1{}_1\,\tilde E_{\a}^1
\\
\nn &&\;  -\frac16\,C^{\ord{1}}_{\a\bd}\, d^1{}_\cssN  
+ \frac12\, C^{\ord{2} 1}_{\a\bd\  \ r}\,
d^r{}_\cssN
+ \frac12\,C^{\ord{3} r}_{\a\bd\ \ \cssN}\, d^1{}_r
\ , 
\end{eqnarray} 
where     $N^{ij}_{\alpha\beta}$      is     given     in~(\ref{e:N}),
$\bar N_{\ad\bd \,ij}$ is its complex conjugate and where
\begin{eqnarray}
  \label{e:C1}
\nn C^{\ord{1}}_{\b\ad}&=&
\begin{cases}
  \bar\chi_{\dot\alpha}^{            rst}\chi_{\beta\,            rst}
  &\textrm{for}~\cN=8~\textrm{and}~\cN\leq 4\\
\bar\chi_{\dot\alpha}^{   rst}\chi_{\beta\,   rst}\,   -   \frac{2}{5}
\chi_\beta^{rstuv} \bar \chi_{\ad\,  rstuv} -4 \chi_\beta^{1\cssN rst}
\bar \chi_{\ad 1\cssN rst}& \textrm{for}~\cN=5,6
\end{cases}\\
C^{\ord{2}1}_{\b\ad\  \ r}&=&
\begin{cases}
   \bar\chi_{\dot\alpha}^{1st}\chi_{\beta\,rst}&\textrm{for}~\cN=8~\textrm{and}~\cN\leq 4\\
\bar\chi_{\dot\alpha}^{1st}\chi_{\beta\, rst} + \frac13 \chi_\beta^{1stuv} \bar \chi_{\ad\,rstuv}& \textrm{for}~\cN=5,6
\end{cases}\\
\nn C^{\ord{3} r}_{\b\ad\ \ \cssN}&=&
\begin{cases}
  \bar    \chi_{\dot\alpha}^{rst}    \chi_{\beta\,    \cssN   st}    &
  \textrm{for}~\cN=8~\textrm{and}~\cN\leq 4\\
\bar \chi_{\dot\alpha}^{rst} \chi_{\beta\, \cssN st} + \frac13 \chi_\beta^{rstuv} \bar \chi_{\ad\,\cssN stuv}& \textrm{for}~\cN=5,6 \,.
\end{cases}
\end{eqnarray}

We define a Grassmann-, or G-analytic, field on $M_H$ to be one that is annihilated by $D_\ha$ and a harmonic-, or H-analytic, field to be one that is annihilated by $d^1{}_\cssN,d^1{}_r,d^r{}_\cssN$. Since the coset $F$ is a complex compact manifold, it follows that H-analytic fields, which are analytic in the usual sense on $F$, have short harmonic expansions, and since we are on-shell, our superfields will be of this type. For G-analyticity, we note that the derivatives $D_\ha$ will contain connection terms with respect to the structure group $SL(2,\IC)\times H$, and that there may be restrictions on the representations under which they can transform. Indeed, Lorentz scalar G-analytic fields can only be charged with respect to a certain $U(1)$ subgroup of $H$ in such a way that they carry sets of  indices with the same number of upper $1$ and lower $\csN$ indices and no others. This restriction follows from the fact that the anticommutator of $D_\ha$ and $D_\hb$ will involve the curvature $R_{\ha\hb}$ with values in $\gh$. The proof that this restriction is required again requires details of the $U(\csN)$ curvature tensor. We have
\be \label{e:Rzero3} R^{1\, 1\ 1}_{\a\b,\ \, 1} =R^{1\, 1\ \cssN}_{\a\b,\ \, \cssN}=0\ ,
\ee 
and similarly for two dotted indices, while
\be \label{e:RB1} R_{\a \bd \cssN,\ \, 1}^{1\ \ \ \ 1}=R_{\a \bd \cssN,\ \,\cssN}^{1\ \ \ \ \cssN}=-\frac{1}{2}  H_{\a\bd\cssN}^1\ .
\ee 
It is the latter equation that shows the need to match the upper $1$ and lower $\csN$ indices for G-analytic fields. The tensor appearing on the right-hand side of this equation will play a key role in the following so we give it its own name,
\be
B_{\a\bd}:=2H^1_{\a\bd \cssN}\ .
\ee
Explicitly, we have
\be 
  \label{e:Bdef}
  B_{\alpha\bd}=
  \begin{cases}
\bar\chi_{\dot\beta}^{1ij}\chi_{\alpha\, \cssN ij}
& \mbox{for}~\cN=4,5, 8\cr
\bar\chi_{\dot\beta}^{1ij}\chi_{\alpha\, 6 ij} +\frac{1}{3} \chi_{\alpha}^{
1ijkl}\bar\chi_{\dot\beta\, 6 ijkkl}
&
 \mbox{for}~\cN= 6 \ ,\cr
  \end{cases}    
\ee 
where $i,j,k,l$ are $(S)U(\csN)$ indices.

The field $B_{\a\ad}$ is also G-analytic, and since it also carries Lorentz indices there is an additional integrability condition that it has to satisfy, namely
\be
R_{\ha\hb, \c}{}^\e B_{\e\bd}-R_{\ha\hb, \bd}{}^\ed B_{\c\ed}=0\ .
\ee
The fact that this is true follows from the explicit forms for these curvatures,
\be \label{e:RB2}
R_{\alpha}^1{}_{\bd \cssN}{}_{,\gamma}{}^\delta=\frac12\,        \delta^\delta_\alpha\,
B_{\gamma\bd}-\frac14\, \delta^\delta_\gamma \, B_{\alpha\bd}\ ,\qquad 
R_{\alpha}^1{}_{\bd \cssN}{}_{,\cd}{}^\dd=\frac12\,        \delta^\dd_\bd\,
B_{\alpha\cd}-\frac14\, \delta^\dd_\cd \, B_{\alpha\bd}\ ,
\ee 
and 
\be \label{e:RB3} R_\alpha^1{}_\beta^1{}_{, \gamma}{}^\delta = R_\alpha^1{}_\beta^1{}_{, \dot{\gamma}}{}^\dd = R_{\ad \cssN\bd\cssN, \gamma}{}^\delta=  R_{\ad \cssN\bd\cssN, \dot{\gamma}}{}^\dd  = 0 \ . \ee

\subsection{Normal coordinates}

We are now going to evaluate on-shell the volume of $\cN$-extended superspace. In principle, one could do this explicitly but it would be extremely tedious. Instead, we shall make use of the normal coordinate method, introduced for superspace in ~\cite{Ogievetsky:1980de} and further developed in~\cite{McArthurFM,Gates:1997ag}, to rewrite the volume integral as an integral over $4(\csN-1)$ odd coordinates using the harmonic superspace formalism. Our discussion follows that of \cite{KuzenkoRY}  where the volume of $\cN=2$ superspace was reduced to a chiral integral by this method. 

Although the conditions for the existence of normal coordinates~\cite{McArthurFM}
\begin{eqnarray}
\label{e:NormalCoordinates}
\zeta^{\hat{A}}:=\{\zeta^{\alpha}:=\delta^\alpha_\mu  \theta^{\mu}_i u^i{}_1\,
,\bar\zeta^{\dot\alpha}:= \delta^{\ad}_{\dot\mu} u^\cssN{}_i \, \bar\theta^{\dot\mu\, i} ,z^r{}_1,z^\cssN{}_r,z^\cssN{}_1 \}
\end{eqnarray}
associated to the vector fields~(\ref{e:Ehat}) are satisfied, one must take into account the fact that these vector fields are only defined on the complexified tangent space, and one must therefore consider the associated normal coordinate expansion as a `holomorphic' expansion in complex coordinates, rather than describing strictly the expansion in coordinates parametrising geodesics normal to a submanifold.

Nevertheless,  the  conditions  assumed  in~\cite{KuzenkoRY}  for  the
expansion in normal coordinates  of the superspace vielbein Berezinian
are satisfied  (since the vector fields $\hat{E}_{\hat{A}}$  are in involution
and span  a representation of  the structure group), and accordingly the harmonic
superspace  vielbein Berezinian  $\tilde E  = E\times  V(t)$, where  $V(t)$  is the
determinant of the vielbein~(\ref{e:VIJ}) over $\mathbb F_{1,1}(\cN)$,
satisfies the flow equation 
\be  \zeta^{\hat{A}}  \partial_{ \hat{A}}  \ln  \tilde  E  =
(-1)^{\HSSA} \bigl( \Omega_{\HSSA\, \hat{B}}{}^{\HSSA} \zeta^{\hat{B}}
-  \zeta^{\hat{B}} T_{\hat{B}\HSSA}{}^{\HSSA} \bigr)  + (-1)^{\hat{M}}
\delta_{\hat{M}}^{\hat{A}}     \bigl(     E_{\hat{A}}{}^{\hat{M}}    -
\delta^{\hat{M}}_{\hat{A}} \bigr) \ , \label{Hflow} \ee 
where we have introduced the notation $\partial_{\hat{A}}:=\partial/\partial_{\zeta^{\hat{A}}}$.
One computes that 
\be  (-1)^{\underline A} T_{\hat{B}\HSSA}{}^{\HSSA} = 0 \ . \label{TracelessTorsion}  \ee 
Note, moreover, that the same formula applies to the flag manifold $ \IF_{1,1}(\csN)$ itself for the expansion of the vielbein determinant $V(t)$ in terms of the normal coordinates $z^R := (z^r{}_1,z^\cssN{}_r,z^\cssN{}_1)$. Since $V(t)$ does not depend on the fermionic variables by construction, one can decompose 
\be \zeta^{\hat{A}}  \partial_{ \hat{A}}  \ln  \tilde E  =
  z^R \partial_{R} \ln V(t) + \zeta^{\hat{A}} \partial_{{\hat{A}}}
  \ln E \ , \ee 
and, removing the pure harmonic component of equation~(\ref{Hflow}), one computes that the superspace vielbein Berezinian $E(x,\theta)$ satisfies the flow equation 
\begin{eqnarray}\zeta^{\hat{A}} \partial_{ \hat{A}} \ln E &=& - \bigl(
  E_\beta^{1\,  M}  \Omega_{M\,  \alpha}{}^\beta  +   E_\alpha^{1\,  M}
   \Omega_{M}{}^1{}_1 \bigr) \zeta^\alpha  + \bigl(   E_{\bd \cssN}{}^{M} \Omega_{M\, \ad}{}^\bd +  E_{\ad \cssN}{}^{M}  \Omega_{M}{}^\cssN{}_\cssN \bigr) \zeta^\ad \nn\\ && \;
-  \delta_{{\mu}}^{\alpha} \bigl( E_{\alpha}^1{}\;^{\mu}_1 -
\delta^{{\mu}}_{\alpha}   \bigr)   -   \delta_{{\dot\mu}}^{\dot\alpha}
\bigl(     E_{\dot\alpha    \cssN}{}^{\dot\mu    \cssN}    -
\delta^{{\dot\mu}}_{\dot\alpha}                \bigr)   \ . \label{Gflow} \end{eqnarray}
The right-hand side is left invariant by the
derivatives  $d^1{}_r, d^r{}_\cssN,d^1{}_\cssN$.  To  show this, we
first note that, thanks to~\eqref{e:Rzero1},~\eqref{e:Rzero2}
and~\eqref{e:Rzero3},  the normal-coordinate gauge condition 
\be  \bigl( \Omega_{\ha}{}_{\HSSA}{}^{\HSSB}
  \bigr)\big|_{\zeta=0} = 0 \ee 
extends to arbitrary $\zeta^{\hat{A}}$ for the components
\be 
\Omega_{\ha}{}^1{}_r = \Omega_{\ha}{}^{r}{}_\cssN = \Omega_{\ha}{}^{1}{}_\cssN =
0\ . \ee 
It follows that one can neglect the harmonic components in $\Omega_{M}{}^1{}_1$ and $ \Omega_{M}{}^\cssN{}_\cssN$ when checking that the right-hand side of~(\ref{Gflow}) is left invariant by $d^1{}_r, d^r{}_\cssN,d^1{}_\cssN$. We conclude that the super-vielbein Berezinian $E(x,\theta)$ does not depend on the coordinates $z^r{}_1,z^\cssN{}_r,z^\cssN{}_1$ and thus one can consider consistently its normal-coordinate expansion in terms of the Grassmann variables $\zeta^{\hat{\alpha}} : = ( \zeta^\alpha, \zeta^\ad)$ alone, \ie
\be   \zeta^{\hat{\alpha}}  \partial_{\hat{\alpha}   }  \ln   E   =  -
\Omega_{\hat{\beta}\,                     \hat{\alpha}}{}^{\hat{\beta}}
\zeta^{\hat{\alpha}}    -    \delta_{{\hat\mu}}^{\hat\alpha}    \bigl(
E_{\hat{\alpha}}{}^{\hat{\mu}}    -    \delta_{{\hat\mu}}^{\hat\alpha}
\bigr) \ , \ee
At this point, the computation goes exactly as in~\cite{KuzenkoRY}, and one deduces that the flow equation can be rewritten as\footnote{The summation convention is such that for fermion bilinears one has $\phi^{\hat\alpha} \psi_{\hat\alpha}= \phi^{\alpha}_1\psi^1_{\alpha}+\bar\phi_\ad^\csssN\bar\psi^\ad_\csssN$.}
\begin{eqnarray}
  \label{e:flow}
\nn  \zeta^{\hat\alpha} \partial_{{\hat{\alpha}} } \ln E &=&\frac13\,
  R_{\hat\gamma\hat\alpha\hat\beta}{}^{\hat\gamma}|_{\zeta=0}\zeta^{\hat\alpha}\zeta^{\hat\beta}+\frac{1}{45}
R_{\hat\eta\hat\alpha\hat\beta}{}^{\hat\rho}R_{\hat\rho\hat\gamma\hat\delta}{}^{\hat\eta}|_{\zeta=0}\zeta^{\hat\alpha}\zeta^{\hat\beta}\zeta^{\hat\gamma}\zeta^{\hat\delta}\\
&&+\frac{5}{12}
D_{\hat\gamma}R_{\hat\delta\hat\alpha\hat\beta}{}^{\hat\delta}|_{\zeta=0}\zeta^{\hat\alpha}\zeta^{\hat\beta}\zeta^{\hat\gamma}-\frac{3}{40} D_{\hat\alpha}D_{\hat\beta}R_{\hat\rho\hat\gamma\hat\delta}{}^{\hat\rho}|_{\zeta=0}\zeta^{\hat\alpha}\zeta^{\hat\beta}\zeta^{\hat\gamma}\zeta^{\hat\delta}\ .
\end{eqnarray}
Note  that one  can  consider the  Riemann  tensor to  be  that  of
$M$ (with appropriate harmonic projections), since those of its components  that are torsion components on
$M_H$ do not contribute to this equation. The curvature components appearing in (\ref{e:flow}) are expressible in terms of $B_{\a\ad}$~\eqref{e:Bdef}, as one can see from~\eqref{e:RB1}, \eqref{e:RB2} and~\eqref{e:RB3}. 
The G-analyticity conditions of $B_{\alpha\bd}$\,, \ie $D_\gamma^1B_{\alpha\bd}=\bar D_{\cd\,\cssN}\,B_{\alpha\bd}=0$\,,
imply that the second line in~(\ref{e:flow}) vanishes for all $
\cN$.  Therefore, the flow equation takes the form

\be 
  \label{e:flow2}
  \zeta^{\hat\alpha} \partial_{{\hat{\alpha}} } \ln E  =-\frac13\,
  B_{\alpha\bd}\zeta^{\alpha}\bar\zeta^{\bd}+\frac{1}{18}
B_{\alpha\bd}B_{\alpha\ad}\zeta^{\alpha}\zeta^{\beta}\bar\zeta^{\ad}\bar\zeta^{\bd}\ .
\ee 
Integrating this equation, we conclude that, for all $\cN$, the supervielbein Berezinian has the expansion

\be 
  \label{Vdensity}
  E(\hat x,\zeta,\bar\zeta)=\cE(\hat x)\,\left(1
-\frac16\, B_{\alpha\bd}
\zeta^{\alpha}\zeta^{\bd}\right)\ ,
\ee 
where $\hat{x}$ stands for all the harmonic superspace coordinates aside from $\zeta^{\hat{\alpha}}$.

In the end, we are not forced to consider the expansion of the fibre determinant $V(t)$ in normal coordinates, and so we can avoid dealing with the issue of reality of the ``holomorphic'' expansion in the variables $z^R$. Moreover, the expansion of $ E(\hat x,\zeta,\bar\zeta)$ is manifestly real with respect to the twisted anti-involution \cite{Galperin:1984av,Hartwell:1994rp}
\be ( u^1{}_i )^* = u^i{}_\cssN \ , \quad ( u^\cssN{}_i )^* = - u^i{}_1 \ , \quad ( u^I{}_i )^* = u^i{}_I \ , \ee
preserving G-analyticity, and so one is ensured that the integral is real. 

We conclude that the superspace volume, subject to the vacuum equations of motion, vanishes for all $\cN$: 
\bea
  \cV_\cssN&=&\kappa^{2(\cssN-2)}\, \int d^4x\, d^{4\cssN}\theta\, E(x,\theta) \CR
  &=& \kappa^{2(\cssN-2)}\, \int d\mu_{\scriptscriptstyle    (\cssN,1,1)}\,
  d^4\zeta \left(1
-\frac16\, B_{\alpha\bd}
\zeta^{\alpha}\zeta^{\bd}\right) = 0 \ ,
\eea
where $\kappa^2$ is Newton's constant (in four dimensions) and we have introduced the $1/\csN$-BPS harmonic
measure $d\mu_{\scriptscriptstyle (\cssN,1,1)}$ defined as
\be 
  \label{e:mu811}
  d\mu_{\scriptscriptstyle    (\cssN,1,1)}:=d^4x\, d^{4\cssN-6} t \, d^{2(\cssN-1)}\theta\, d^{2(\cssN-1)}\bar\theta\,\cE(\hat x)V(t)\ .
\ee 
At the  linearised level, this reduces to the  measure discussed
in~\cite{Drummond:2003ex}.  In  the   next  section  we  will discuss  some
properties of this measure.

\section{Full superspace integrals}
\label{sec:volume}

Let us now interpret formula~(\ref{Vdensity}). 
The normal-coordinate expansion  of a generic scalar superfield $\Phi$
(not necessarily of mass dimension 0) is~\cite{KuzenkoRY}
\be 
 \Phi = \exp\left( \zeta^{\hat A} D_{\hat A} \right) \Phi
 \bigg|_{\zeta = 0} \ .
\ee 
However, because $\Phi$ does not depend on the harmonic variables and because the covariant derivatives in the harmonic direction commute with the Grassmann covariant derivatives, this expansion reduces to
\be 
 \Phi = \exp\left( \zeta^{\hat \alpha} D_{\hat \alpha} \right) \Phi
 \bigg|_{\zeta = 0} \ .
\ee 
The expansion of the vielbein Berezinian is such that 
\be 
   \biggl( 1 - \frac{1}{6} B_{\alpha\bd} \zeta^{\alpha} \bar\zeta^{\bd} \biggr)
   \exp\left( \zeta^{\hat{\alpha}}  D_{\hat{\alpha}}\right) \Phi \bigg|_{\zeta
     = 0} =  \frac{1}{2} \Bigl\{ \exp\left( \zeta^{\alpha} D^1_{\alpha}\right)
   ,  \exp\left(   \zeta^{\ad}  \bar D_{\ad \cssN}  \right)   \Bigr\}  \Phi
   \bigg|_{\zeta = 0} 
 \ee 
and so  it plays the  role of a  normal-ordering operator.  It follows
that \bea
 \int  d^4x d^{4\cssN}  \theta \,  E(x,\theta) \,  \Phi&=& \frac{1}{2}
 \int d\mu_{\scriptscriptstyle     (\cssN,1,1)}  \, d^4 \zeta  \, \Bigl\{ \exp\left( \zeta^{\alpha} D^1_{\alpha}\right)
   ,  \exp\left(   \zeta^{\ad}  \bar D_{\ad\cssN}  \right)   \Bigr\}  \Phi
   \bigg|_{\zeta = 0} \CR
\label{e:Dterm}   &=&  \frac{1}{4} \int
 d\mu_{\scriptscriptstyle     (\cssN,1,1)}     \,    \Bigl( (D^1)^2 (\bar D_\cssN)^2
 \, \Phi \Bigr)\Big|_{\zeta=0} \ , 
\eea
where  $  (D^1)^2:=\varepsilon^{\alpha\beta}D^1_{\alpha}  D^1_{\beta}$
and $(\bar D_\cssN)^2:= \varepsilon^{\ad\bd} \bar D_{\ad \cssN}
 \bar D_{\bd\cssN}$
and where we have used the commutation property 
\be  
[(D^1)^2,(\bar D_\cssN)^2]=0\ . \label{CommuteSquare}
\ee 
Therefore, the form of the Berezinian derived in the previous section implies
that any full superspace integral can be rewritten as an integral over
the harmonic measure~(\ref{e:mu811}).
Conversely, using this measure 
 one can define supersymmetric invariants for any G-analytic integrand.
 The  integrand  in~(\ref{e:Dterm})  is  indeed  G-analytic  with  the
 correct $U(1)$-charges.  

We note further that this confirms the vanishing of the full superspace volume, because it can be thought of as the integral of $\Phi=1$ over the full superspace.

In the  following section, we will use the harmonic  measure to
construct non-vanishing supersymmetric duality invariants.

\section{Invariants in extended superspace}
\label{sec:invariants}

The result that the extended superspace volumes all vanish might be considered disturbing, since one expects the existence of a duality invariant of this dimension from the linearised supersymmetry analysis~\cite{Beisert:2010jx}. Nevertheless, we shall see that such invariants do indeed exist as $1/\csN$-BPS integrals.

\subsection{$(\cN-1)$-loop supersymmetric \& duality invariants}
\label{sec:counterterms}

By integrating G-analytic  quartic expression in the fermions over
the harmonic  measures $d\mu_{\scriptscriptstyle (\cssN,1,1)}$\,,  we obtain a set of fully supersymmetric  duality-invariant  integrals 
\be 
  \label{e:In}
  I^{\cssN}:= \kappa^{2(\cssN-2)}\, \int                 d\mu_{\scriptscriptstyle (\cssN,1,1)}                \,
    B_{\alpha\bd}\,B^{\a\bd} \ . 
\ee 
One  can  check that  the  integrand  of  \eqref{e:In} is  the  unique duality-invariant
G-analytic scalar superfield at this dimension for $\cN=4,5,8$. This is also the G-analytic duality-invariant scalar operator of smallest mass dimension. 
 We will show that this  reduces to the quartic invariant $\int  d^{4\cssN} \theta\,( W_{ijkl} \bar W^{ijkl} )^2\sim (\partial^{\cssN-4} C \bar C )^2$ in the linearised approximation.  

For $\cN=6$ there is an additional integral
\be \label{e:In2}
I^{6}_2:=\kappa^8\,  \int  d\mu_{\scriptscriptstyle (6,1,1)}                \,
  \varepsilon^{\alpha\beta}\varepsilon^{\ad\bd}    \biggl(J_{\alpha\bd}{}^{1i}_{6  i} J_{\beta\ad}{}^{1j}_{6  \,j} +\frac{4}{3}\,
J_{\alpha\bd}{}^{1i}_{6\,j} J_{\beta\ad}{}^{1j}_{6 \,i}\biggr)\ ,
\ee 
which we will show to  correspond to an independent combination of $\int  d^{24} \theta\,( W_{ijkl} \bar W^{ijkl} )^2$ and the additional linearised quartic invariant $\int \, d^{24}\theta \; W_{ijkl} \bar W^{klmn} W_{mnpq}\bar W^{pqij}$. These two invariants contribute to the two inequivalent forms of $(\partial^{2} C \bar C)^2$.

These expressions are non-vanishing, fully supersymmetric and
duality-invariant  candidate counterterms that could correspond to   $(\csN-1)$-loop logarithmic divergences in four-dimensional $\cN$-extended
supergravity.

Importantly, these invariants cannot be rewritten as full superspace integrals
because there is no duality-invariant dimension-zero scalar superfield $\Phi$ such
that the  integrand of~(\ref{e:In}) is given by
$(D^1)^2(\bar D_{\cssN})^2\Phi$.  We will see below that such a scalar can be found 
at the linearised order but that it does not extend to the full theory in a duality-invariant way.

\bigskip

\noindent $\bullet$ For $\cN=4,6$ and $\cN=8$, we evaluate the integral in~\eqref{e:In} in the linearised
approximation. First of all, we note that in this approximation the scalar superfield $W_{ijkl}$ satisfies the linear constraints 
\be D_\alpha^p W_{ijkl} = 2\delta^p_{[i} \chi_{\alpha\, jkl]} \ , \quad \bar D_{\ad  p}  W_{ijkl} = \bar \chi_{\ad\, pijkl} \  , \ee
and similarly for its complex conjugate $\bar W^{ijkl}$. For $\cN=8$, $ \bar W^{ijkl} =  \frac{1}{24} \varepsilon^{ijklmnpq} W_{mnpq}$.  As a direct consequence, $W_{ijkl}$ and $\bar W^{ijkl}$ also satisfy the quadratic constraints 
\be D_{\alpha}^1 D_\beta^1 W_{ijkl} = D_{\alpha}^1 D_\beta^1 \bar W^{ijkl} = \bar D_{\ad\cssN} \bar D_{\bd \cssN} W_{ijkl} =\bar D_{\ad\cssN} \bar D_{\bd \cssN} \bar W^{ijkl} = 0 \ .   \ee
 The components  $W_{1rs\cssN}$ and their complex conjugates satisfy in particular
\be D_{\alpha}^1 W_{1rs\cssN} =  \frac12 \chi_{\alpha\, \cssN rs} \ ,
\quad D_{\alpha}^1 \bar W^{1rs\cssN} = 0 \ , \quad \bar D_{\ad \cssN}
W_{1rs\cssN} = 0 \ , \quad \bar D_{\ad \cssN} \bar W^{1rs\cssN} = - \frac12
\bar \chi_\ad^{1rs} \ . \ee
It follows, in the linearised approximation, that for $\cN=4,5$ and 8, one has
\be    (D^1)^2 (\bar D_\cssN)^2 \bigl( W_{1rs\cssN} \bar
W^{1rs\cssN} \bigr)^2 = \frac{1}{4} B_{\alpha\bd} B^{\a\bd} \ . \ee
The integration over the harmonic  variables is done using the measure
$du := d^{4\cssN-6}t\,V(t)$ with respect to which one has the relations
\be
  \label{eq:norm}
  \int du \, 1=1\; ,\quad 
\int du\, u^i{}_1 u^1{}_j=\int du \, u^i{}_\cssN u^\cssN{}_j=\frac{1}{\csN}\delta^i_j\; ,\quad
\int  du \, u^i{}_1  u^\cssN{}_j=0\ ,
\ee
and
\begin{eqnarray}
\nn&&\int du\,  u^{i_1}{}_1  u^{i_2}{}_1   u^1{}_{k_1}  u^1{}_{k_2}
u^{j_1}{}_\cssN u^{j_2}{}_\cssN u^\cssN{}_{l_1} u^\cssN{}_{l_2}
= \frac{4}{ (\csN-1)\csN^2(\csN+2)(\csN+3)} \\
&&\times\Big( (\csN+2)\,\delta^{(i_1}_{(k_1}\delta^{i_2)}_{k_2)}\delta^{(j_1}_{(l_1}\delta^{j_2)}_{l_2)} - 4
\,\delta^{(i_1}_{(k_1}\delta^{i_2)}_{(l_1|}\delta^{(j_1}_{|k_2)}\delta^{j_2)}_{l_2)}
+ \frac{2}{ \csN+1} \,\delta^{(i_1}_{(l_1}\delta^{i_2)}_{l_2)}\delta^{(j_1}_{(k_1}\delta^{j_2)}_{k_2)}\Big)\ .
\end{eqnarray}
Using this result, we find that
\begin{eqnarray}
  \label{e:InF}
I^\cssN&=&      \kappa^{2(\cssN-2)}      \int     d^4 x\,du\,d^{4\cssN} \theta\,
(W_{1rs\cssN}\bar W^{1rs\cssN})^2\\
\nn&=&\frac{\kappa^{2(\cssN-2)}}{
      (\csN^2-1)\csN^2}\, \int d^4x\, d^{4\cssN}\theta\,(I_1+2I_2+I_3)\ ,
\end{eqnarray}
where
\be 
  I_1=(W_{ijkl}\bar W^{ijkl})^2\ , \quad
  I_2=W_{ijkl} \bar W^{ijkm} W_{npqm}\bar W^{npql}\  , \quad
  I_3=W_{ijkl}\bar W^{klmn} W_{mnpq} \bar W^{pqij}\ .
\ee
We have  $I_3=I_1/6$ and $I_2=I_1/4$ for $\cN=4\,\&\,5$, and $I_3=I_1/12$ and
$I_2=I_1/8$ for $\cN=8$.
We conclude that, in the  linearised approximation for $\cN=4,5$ and 8,
$I^\cssN$ evaluates to yield the full superspace integrals analysed in~\cite{Howe:1980th,Beisert:2010jx}:

\begin{eqnarray}
  \label{e:InF2}
    I^\cssN
&=&\kappa^{2(\cssN-2)}\frac{5-\delta_{\cssN,8}}{
      3(\csN^2-1)\csN^2}\,  \int  d^4x\, d^{4\cssN}\theta\,  (W_{ijkl}\bar
    W^{ijkl})^2\\
\nn&\sim&\kappa^{2(\cssN-2)}\int   d^4x\,   (\partial^{\cssN-4}(\bar
C_{\ad\bd\cd\dd} C_{\a\b\c\d})\partial^{\cssN-4}(\bar C^{\ad\bd\cd\dd}
C^{\a\b\c\d})+{\rm s. s. c. })\ .
\end{eqnarray}
As   shown in detail   in Appendix~\ref{sec:Kinematics},  these   linearised
expressions are unique.  Because $\cN$-extended supergravity admits an enhanced  $SU(2,2|\csN)$  superconformal  symmetry in  the  linearised
approximation, one  can use superconformal representation theory
to determine the number of independent integrands defined as functions
of  the  scalar superfields~\cite{Drummond:2003ex,Drummond:2010fp}.  A
$U(\csN)$  scalar  monomial in  $(W\bar W)^{n}$  is  a superconformal  primary
operator of conformal weight $2n$,  and zero R-charge whereas the only
short  such  superconformal   primary  operators  are  necessarily  of
conformal weight $2$ (or zero)~\cite{Drummond:2010fp,Dobrev:1985qv}.
 So it follows that any independent $U(\csN)$ scalar monomial of order
 four in  $W$ gives rise to  a non-trivial superspace  integral in the
 linearised approximation which is not  a total derivative, and which can be
 shown to include $(\partial^{\cssN-4} C \bar C)^2$ type terms.\footnote{The leading $\partial^4 F^4$ term in the analogous $(W_{ij} \bar W^{ij})^2$ integrand in $\cN=4$ abelian super Yang--Mills theory was evaluated  explicitly in \cite{Drummond:2003ex}.} To see this property explicitly in $\cN=8$ supergravity, it is convenient to consider a formulation in $(8,4,4)$ linearised harmonic superspace. We note here that, although this harmonic superspace formulation cannot be extended to the non-linear level, it is perfectly well defined in the linear approximation \cite{Hartwell:1994rp}. Using the linear constraints on $W_{ijkl}$, one computes that 
 \be (D^1)^2  (D^2)^2 (D^3)^3 (D^4)^2 (\bar D_5)^2  (\bar D_6)^2 (\bar
 D_7)^2 (\bar D_8)^2 \; (W_{ijkl} \bar W^{ijkl} )^2 \sim ( \partial^2 W_{1234} )^4 \ , \ee
because $SU(8)$ considerations imply that the result must be quartic in the $(8,4,4)$ G-analytic superfield $W_{1234}$\,; and this expression cannot be a total derivative because $(W \bar W)^2$ is a long primary operator. It is straightforward to check that the contractions of the derivatives are uniquely fixed by Lorentz invariance up to a total derivative. Using the property that the derivatives commute with integration over the fermionic variables, together with the fact that $(W_{1234})^4$ integrates in $(8,4,4)$ superspace to yield the linearised $(C \bar C)^2$ invariant \cite{Drummond:2003ex}, one concludes that
\bea \int d^4 x\, d^{32} \theta (W_{ijkl} \bar W^{ijkl} )^2 &\sim& \int d\mu_{\scriptscriptstyle (8,4,4)} ( \partial^2 W_{1234} )^4 \CR
 &\sim& \int d^4 x  \bigl( ( \partial^2  C \partial^2 \bar C )^2 + \mbox{s.s.c.} \bigr)  \ , \eea
 which clearly coincides with the invariant exhibited in~\cite{FreedmanR4}.

\medskip
\noindent$\bullet$ For the $\cN=6$ case, one must consider in addition the components $W_{rstu}$ and their complex conjugates, which satisfy
\be D_{\alpha}^1 W_{rstu} = 0  \ , \quad D_{\alpha}^1 \bar W^{rstu} =
\chi_\alpha^{1rstu}  \  ,  \quad  \bar  D_{\ad  6}  W_{rstu}  =  \bar
\chi_{\ad\, 6rstu} \ , \quad \bar D_{\ad 6} \bar W^{rstu} =0 \ .\ee
Note that $W_{rstu}$ with $2\leq r,s,t,u\leq \csN-1$ vanishes identically for $\cN < 6$, and is equal
to $\frac{1}{2} \varepsilon_{rstuvw} \bar W^{1vw8}$ for $\cN=8$.  In
$\cN=6$, one has in the linearised approximation
\be    (D^1)^2 (\bar D_6)^2 \biggl( W_{1rs6} \bar W^{1rs6} +
\frac{1}{12}   W_{rstu}   \bar   W^{rstu}  \biggr)^2   =   \frac{1}{4}
B_{\alpha\bd} B^{\a\bd} \ .  \ee
The invariant (\ref{e:In}) evaluates to give
\begin{eqnarray} 
  I^6_1 := I^6 &=& \kappa^{8}\, \int d\mu_{(6,1,1)}  \Bigl( 4  W_{1rs6} \bar
W^{1rs6} + \frac{1}{3} W_{rstu} \bar W^{rstu} \Bigr)^2\\ 
\nn              &=             &\frac{\kappa^{8}}{945}\,\int             d^4x\,
d^{24}\theta\, \Bigl( 23 ( W_{ijkl} \bar W^{ijkl} )^2 + 12 W_{ijkl}
\bar W^{klpq} W_{pqmn} \bar W^{mnij} \Bigr )\ ,
\end{eqnarray}
while the invariant (\ref{e:In2}) evaluates to yield
\begin{eqnarray}
 I_2^6&=&\kappa^{8}\, \int d\mu_{(6,1,1)} \Bigl( ( W_{1rs6} \bar W^{1rs6})^2 + \frac{4}{3}
W_{1rs6} W_{1tu6} \bar W^{1tr6} \bar W^{1su6} \Bigr)^2\\
\nn  &=&\frac{\kappa^{8}}{30240}\,\int d^4x\,  d^{24}\theta\, \Bigl(
23 ( W_{ijkl}\bar W^{ijkl} )^2
-58 W_{ijkl}\bar W^{klpq} W_{pqmn}\bar W^{mnij} \Bigr ) \, .
\end{eqnarray}
 These two invariants clearly define the supersymmetrisation of two different combinations of the two linearised independent
 $(\partial \bar C \partial C)^2$  structures that  exist  for  $\cN=6$ (see 
 Appendix~\ref{sec:Kinematics} for details).
Since pure $\cN=6$ supergravity  is   a strict   truncation  of   $\cN=8$  theory,   the
four-graviton amplitudes are different in these theories.

\subsection{ $\nabla^{2(\cN-5)}R^4$ invariants}

Using the $(\csN,1,1)$-measure, by integrating  $G$-analytic functions
of   the   scalar   fields   generalising  the   ones   given in
\cite{Drummond:2003ex}, we can construct nonlinear versions
of the $1/\csN$-BPS invariants of general structure  $\nabla^{2(\cssN-5)}R^4$. These will be invariant under supersymmetry and the the corresponding R-symmetry groups $K$, but not under the continuous duality symmetries $G$ as were the $\nabla^{2(\cN-4)}R^4$ invariants of the last section.

\noindent$\bullet$ For $\cN=8$, let us define the superfield $( SU(8) / \IZ_2)\backslash E_{7(7)}$
representative in the fundamental ${\bf 56}$ representation decomposed
as ${\bf 28} \oplus \overline{\bf 28}$ of $SU(8)$

\be\cV :=  \left( \begin{array}{cc}\; U_{ij}{}^{\gI\gJ} \; & \; V_{ij\gK\gL} \; \\ \; \bar V^{kl\gI\gJ} \; & \; \bar  U^{kl}{}_{\gK\gL} \; \end{array}\right)\ ,  \ee
where $\csI, \, \csJ \dots$ stand for the rigid $SU(8)$ indices while the $i,\, j,\, $ indices stand for local $SU(8)$ indices as used throughout this paper. The derivative $D_\alpha^i$ acts on these superfields as follows 
\bea D_\alpha^k U_{ij}{}^{\gI\gJ} &=& 2 \delta^k_{[i} \chi_{\alpha\, j]pq}\bar  V^{pq\gI\gJ} \ , \quad D_\alpha^k \bar U^{ij}{}_{\gI\gJ} = \frac{1}{12} \varepsilon^{ijklmnpq} \chi_{\alpha\, lmn} V_{pq\gI\gJ}  \ ,  \CR
D_\alpha^k V_{ij\gI\gJ} &=& 2 \delta^k_{[i} \chi_{\alpha\, j]pq} \bar U^{pq}{}_{\gI\gJ} \ , \quad D_\alpha^k \bar V^{ij\gI\gJ} = \frac{1}{12} \varepsilon^{ijklmnpq} \chi_{\alpha\, lmn} U_{pq}{}^{\gI\gJ} \ ,  \eea
and similarly for $\bar D_{\ad  i}$ by complex  conjugation. It follows that
the    superfields    $U_{8r}{}^{\gI\gJ}$,    $V_{8r\gI\gJ}$,    $
\bar U^{1r}{}_{\gI\gJ}$ and  $\bar V^{1r\gI\gJ}$ are  all G-analytic. There
are {\it a  priori} several combinations  of these superfields that  are of
the right $U(1)$ weight and that are  left invariant under the rigid $SU(8)$
symmetry,  but  we are going to see that they are all equivalent because of $E_{7(7)}$  identities, consistently with the property that there is a unique $SU(8)$-invariant G-analytic function of the scalar superfield in the linearised approximation.  A  first  set  of
conditions arises from the fact that~\cite{de Wit:1982ig}
\be 
  \cV^{-1}=
  \begin{pmatrix}\; 
    \bar U^{ij}{}_{\gI\gJ} \;  &   -V_{kl\gI\gJ}  \;  \\  -\bar  V^{ij\gK\gL}  \; &  \;
     U_{kl}{}^{\gK\gL} \; 
  \end{pmatrix}
\ .
\ee 
This implies that the G-analytic superfields satisfy
\be U_{8i}{}^{\gI\gJ} \bar U^{1j}{}_{\gI\gJ} = V_{8i\gI\gJ} \bar V^{1j \gI\gJ} \ , \quad U_{8i}{}^{\gI\gJ} V_{8j\gI\gJ} = U_{8j}{}^{\gI\gJ} V_{8i\gI\gJ} \ ,\quad  \bar U^{1i}{}_{\gI\gJ} \bar V^{1j\gI\gJ} = \bar U^{1j}{}_{\gI\gJ} \bar V^{1i\gI\gJ} \ . \ee

Using the fact that, for any element $X$ of the complex Lie algebra $\mathfrak{e}_7$, $\cV^{-1} X
\cV$ is also an element of $\mathfrak{e}_7$, one deduces further identities satisfied by $U_{ij}{}^{\gI\gJ}$ and $V_{ij\gI\gJ}$~\cite{de Wit:1982ig}. In particular, taking the $\mathfrak{sl}(8,\mathbb{C})\subset \mathfrak{e}_7$ element 
\be X :=  \left( \begin{array}{cc}\; 2 \delta_{[i}^{[k} \, u^1{}_{j]} u^{l]}{}_8  \; & \; 0 \; \\ \; 0 \; & \;- 2 \delta_{[k}^{[i}\,  u^1{}_{l]} u^{j]}{}_8    \; \end{array}\right)\ ,  \ee
one obtains 
\bea 
U_{8i}{}^{\gI\gJ} \bar U^{1i}{}_{\gI\gJ} &=& V_{8i\gI\gJ} \bar  V^{1i\gI\gJ} = 0 \ , \CR
U_{8i}{}^{\gI\gJ} \bar U^{1i}{}_{\gK\gL} + V_{8i\gK\gL} \bar  V^{1i\gI\gJ} &=& \frac{2}{3} \delta^{[\gI}_{[\gK} \Bigl(   U_{8i}{}^{\gJ]\gP} \bar U^{1i}{}_{\gL]\gP} + V_{8i\gL]\gP} \bar  V^{1i\gJ]\gP} \Bigr) \ , \\
\nn \bar U^{1i}{}_{\gI\gJ} V_{8i\gK\gL} + \bar U^{1i}{}_{\gK\gL} V_{8i\gI\gJ} &=&- \frac{1}{12} \varepsilon_{\gI\gJ\gK\gL\gM\gN\gP\gQ} U_{8i}{}^{\gM\gN} \bar V^{1i\gP\gQ} \ . 
\eea 
Using the converse, \ie the fact that for any element $Y$ of the complex Lie algebra  $\mathfrak{sl}(8,\mathbb{C}) \subset \mathfrak{e}_7$, $\cV Y
\cV^{-1}$ is an element of $\mathfrak{e}_7$, one obtains similarly 
\bea U_{8i}{}^{\gI\gK} V_{8j\gJ\gK} +  U_{8j}{}^{\gI\gK} V_{8i\gJ\gK} &=& \frac{1}{4} \delta^\gI_\gJ  U_{8i}{}^{\gK\gL} V_{8j\gK\gL} \ , \\
 U_{8s}{}^{\gI\gK} \bar U^{1r}{}_{\gJ\gK} + V_{8s\gJ\gK} \bar V^{1r\gI\gK} &=& \frac{1}{4} \delta_\gJ^\gI V_{8s\gK\gL} \bar V^{1r\gK\gL} + \frac{1}{6} \delta_s^r \bigl(  U_{8i}{}^{\gI\gK} \bar U^{1i}{}_{\gJ\gK} + V_{8i\gJ\gK} \bar V^{1i\gI\gK} \bigr) \ .\nn \eea
Using these identities, one shows that all G-analytic $SU(8)$ invariant functions of the scalar superfields are determined as functions of one single expression which reproduces the unique  $1/8$ BPS integrand defined in~\cite{Drummond:2003ex}  in the quartic approximation, {\it viz.}
\be  \label{e:F1188}
 {\cal F}^{11}_{88}(\cV) :=  u^1{}_i u^1{}_j  u^k{}_8 u^l{}_8  \, \bar
V^{im\gI\gJ}       \bar        V^{jn\gK\gL}            V_{km\gK\gL}
V_{ln\gI\gJ} \ ,
 \ee 
so that 
\be 
 \int  d\mu_{\scriptscriptstyle  (8,1,1)}  \; {\cal  F}^{11}_{88}(\cV)
 \sim \int  d^4x \,  e \biggl( f^8_{6}(\phi) \nabla^3 R^2 \cdot  \nabla^3 R^2   + {\rm s. s. c. }
 \biggr) \ ,
\ee 
where $\nabla^k R^2$ is the rank $k+4$ symmetric traceless tensor obtained by acting with $k$ covariant derivatives on the Bel--Robinson tensor, and $f^8_6(\phi)$ is the (appropriately normalised) $SU(8)$ invariant function of the 70 scalar fields discussed in \cite{Bossard:2010bd,Beisert:2010jx}. This provides a nonlinear supersymmetric $SU(8)$, but not
$E_{7(7)}$, invariant form for the
1/8-BPS coupling $(\nabla^3R^2)^2$ in $\cN=8$ supergravity.

\medskip

\noindent$\bullet$ For $\cN=6$, we define the superfield $U(6)\backslash SO^*(12)$ representative in the vector representation ${\bf 12}$ decomposed as the ${\bf 6}^\ord{-1} \oplus \overline{\bf 6}^\ord{1}$ of $U(6)$ 
\be\cV :=  \left( \begin{array}{cc}\; U_{i}{}^{\gI} \; & \; V_{i\gJ} \; \\ \; -\bar V^{j\gI} \; & \; \bar  U^{j}{}_{\gJ} \; \end{array}\right) \ , \ee
similarly as for $\cN=8$. In this case, it is enough to use the property that $\cV$ preserves the K{\"a}hler metric 
\be \cG := \left( \begin{array}{cc}\; 0 \; & \; \delta^\gJ_\gK \; \\ \; \delta^\gL_\gI \; & \;  0 \; \end{array}\right) \ , \ee
\ie  $\cV \cG \cV^{T} = \cG$, to find that there is a unique G-analytic superfield of the correct $U(1)$ weight left invariant by the rigid $U(6)$, \ie
\be 
 {\cal  F}^{11}_{66}(\cV) :=  u^1{}_i u^1{}_j  u^k{}_6  u^l{}_6 \,
 \bar V^{i\gI} \bar V^{j\gJ} V_{k\gI} V_{l\gJ} \ .  
\ee 
The resulting integral is of the form 
\be 
 \int d\mu_{\scriptscriptstyle (6,1,1)} \; {\cal F}^{11}_{66}(\cV) \sim   \int d^4x \,e \biggl(f^6_4(\phi)  \nabla  R^2 \cdot \nabla  R^2 + {\rm s. s. c. } \biggr) \ . \ee 
This provides a nonlinear supersymmetric $U(6)$, but not
$SO^*(12)$, invariant for the
$1/6$-BPS coupling $(\nabla R^2)^2$ in $\cN=6$ supergravity.  

\medskip

\noindent$\bullet$ For $\cN=5$, we define the superfield $U(5)\backslash SU(5,1)$ representative in the fundamental representation ${\bf 6}$ decomposed as the ${\bf 1}^\ord{5} \oplus {\bf 5}^\ord{-1}$ of $U(5)$ 
\be\cV :=  \left( \begin{array}{cc}\; U  \; & \; V_{\gI} \; \\ \;  V^{i} \; & \;   U^{i}{}_{\gI} \; \end{array}\right) \ . \ee
In the same way as above, the unique G-analytic superfield of the right $U(1)$ weight that is left invariant by the rigid $U(5)$ is 
\be 
 {\cal F}^{11}_{55}(\cV)  := u^1{}_i  u^1{}_j u^k{}_5 u^l{}_5  \, 
 V^{i}  V^{j} \bar V_{k} \bar V_{l} \ . \ee 
The resulting integral is of the form 
\be  \int d\mu_{\scriptscriptstyle (5,1,1)} \; {\cal F}^{11}_{55}(\cV) \sim \int d^4x \, e \bigl( f^5_3(\phi) R^2 \cdot R^2  + {\rm s. s. c. } \bigr) \  . \ee 
This provides a nonlinear supersymmetric $U(5)$, but not
$U(5,1)$, invariant for the
$1/5$-BPS coupling $R^4$ in $\cN=5$ supergravity.

\subsection{Duality-invariant full-superspace integrals} 
\label{sec:Dterms}

The vanishing of the superspace volume implies that the first duality-invariant full superspace integrals available as invariant candidate counterterms will start from the $\cN$-loop order for $\cN$-extended supergravity. 

For the $\cN=8$ case, the candidate counterterm contributing to four-point amplitudes is the
invariant discussed
in~\cite{Howe:1980th,Kallosh:1980fi} 
\be\label{e:chi4}
I_{(\chi\bar\chi)^2}:=\kappa^{14}\, \int d^4x\, d^{32}\theta\,E(x,\theta)\,  \varepsilon^{\alpha\beta}\varepsilon^{\ad\bd}
\chi_{\alpha\, ijk}\bar\chi_\ad^{ijk}\chi_{\beta\,mnp}\bar\chi_\bd^{mnp} \ .
\ee
It can be computed to give rise to a $(\partial^5 C\bar C)^2$ contribution in the linearised approximation,
\be I_{(\chi\bar\chi)^2} \sim \kappa^{14}\,\int d^4x\,e \bigl(  ( \nabla^5 R^2 )^2 +{\rm s. s. c. } \bigr)  \; . \ee

At the same dimension, there are also the duality invariants
\be\label{e:chi4bis}
I_{\chi^2\bar\chi^2}:=\kappa^{14}\, \int d^4x\, d^{32}\theta\, E(x,\theta)\,\varepsilon^{\alpha\beta}\varepsilon^{\ad\bd}
\chi_{\alpha\, ijm}\bar\chi_\ad^{ijn}\chi_{\beta\,
pqn}\bar\chi_\bd^{pqm} 
\ee
and
\be
\label{e:ccF}
I_{\chi^2 M}:=\kappa^{14}\, \int d^4x\, d^{32}\theta\,E(x,\theta)\, \varepsilon^{\alpha\gamma} \varepsilon^{\delta\beta}\,\varepsilon^{ijklmnpq}
\chi_{\alpha\, ijk}\chi_{\beta\,lmn}M_{\gamma\delta\, pq}\ , 
\ee
where $M_{\alpha\beta\,ij}$ is the dimension-one superfield for the vector field-strengths, {\it viz}
\be \label{e:Fdef}
F_{\alpha\beta,\dot\alpha\dot\beta\,
ij}=-i\varepsilon_{\ad\bd}
\,M_{\alpha\beta\,ij}+\frac{i}{72}\varepsilon_{\alpha\beta}\, \varepsilon_{ijklmnpq}\bar\chi^{klm}_{\ad}\bar\chi^{npq}_{\bd}\ .
\ee 
Using the relation
\begin{eqnarray}\label{e:Identity}  \varepsilon^{\alpha\beta} \varepsilon^{\gamma\delta}
\, \varepsilon^{jklmnpqr}\,
D_\alpha^i \Bigl( \chi_{\gamma\, ijk} \chi_{\beta\, lmn} \chi_{\delta\, pqr} \Bigr) &=& 9 \varepsilon^{\alpha\beta} \varepsilon^{\gamma\delta}  \varepsilon^{ijklmnpq}  M_{\alpha\gamma\, ij} \chi_{\beta\, klm} \chi_{\delta\, npq} \\
&&  - 90   \varepsilon^{\alpha\beta} \varepsilon^{\ad\bd} \chi_{\alpha\, i[jk} \chi_{\beta\, lmn]} \bar \chi_\ad^{i[jk} \bar \chi_\bd^{lmn]} \nn
 \end{eqnarray}
 and (\ref{TracelessTorsion}), one shows that 
\be 
\label{e:ccF2}
I_{\chi^2 M}= 10 \kappa^{14}\, \int d^4x\, d^{32}\theta\,E(x,\theta)\,\varepsilon^{\alpha\beta} \varepsilon^{\ad\bd} \chi_{\alpha\, i[jk} \chi_{\beta\, lmn]} \bar \chi_\ad^{i[jk} \bar \chi_\bd^{lmn]} \ ,
\ee 
because the difference is the superspace integral of a total superspace derivative. We conclude that at mass dimension 18 there are only two nonlinear
supersymmetric duality invariants. These invariants are fully
$E_{7(7)}$ invariant because they are constructed from a full superspace
integral of the superfield entering in the superspace torsion.
They are independent as can easily be seen from the
inequivalent $SU(8)$ structures in~(\ref{e:chi4})
and~(\ref{e:ccF2}).

Since at the linearised order there is only one kinematic structure
$(\partial^5 C\bar C)^2$ contributing to the 4-point amplitude \cite{Beisert:2010jx}, one expects that the second invariant $I_{\chi^2 M}$ will only start contributing at 8-loop order from the five-point amplitude 
\be I_{\chi^2 M} \sim\kappa^{14}\,\int d^4x\,e ( \nabla^8 R^5 +{\rm s. s. c. }) \ . \ee  
This can be proved using the analysis in~\cite{Beisert:2010jx} which states that the superconformal representation theory of $SU(2,2|8)$ implies that there is only one linearised invariant of this dimension that contributes first at four points, and only one complex (two real) linearised invariant that contributes first at five points. They are the only invariants of this dimension that are left invariant by a shift of the scalar fields in the linearised approximation. However, the parity-odd linearised five-point invariant does not extend at the non-linear level to a duality-invariant full superspace integral, because the imaginary part of $I_{\chi^2 M}$ is the integral of a total derivative and thus vanishes. It is possible that there exists a duality-invariant parity-odd invariant which would be defined as the $(8,1,1)$ harmonic superspace integral of a G-analytic superfield of mass-dimension 4. We will not investigate this possibility further because such an invariant would be ruled out as a possible counterterm by the odd parity.

To understand why the invariant associated to the cubic integrand~(\ref{e:ccF}) indeed starts contributing only from five points, it is relevant to compare it to the linearised Konishi operator $W_{ijkl} \bar W^{ijkl}$. They both satisfy the quadratic constraint
\be \varepsilon^{\alpha\beta} D_\alpha^i D_\beta^j L = \varepsilon^{\ad\bd} \bar D_{\ad i} \bar D_{\bd j} L = 0 \ , \label{quadraLin} \ee
in  the   linearised  approximation \cite{Drummond:2003ex}.  Their   superspace  integrals 
therefore   vanish  in   the  linearised   approximation.  However,
computing the G-analytic descendent of the na{\"\i}ve nonlinear equivalent
of  the  Konishi  operator,  \ie  $V_{ij\gI\gJ}  \bar  V^{ij\gI\gJ}$,
according to  formula~(\ref{e:Dterm}), one obtains that  $(D^1)^2 (\bar D_8)^2
V_{ij\gI\gJ} V^{ij\gI\gJ}$ is quartic  in fields in the linearised
approximation, and the corresponding terms can be identified with $(D^1)^2 (\bar D_8)^2 (W \bar W)^2$ in  this approximation. We conclude  therefore that
the  existence  of  the  $B_{\a\ad}$  term  in  the  normal-coordinate
expansion  of  the supervielbein  Berezinian  has  the  effect that  a  
superfield  $L$  satisfying the quadratic  constraint~(\ref{quadraLin})
 in  the  linearised approximation, without being a total derivative at the non-linear level, is  effectively  equivalent to  the
operator $(W \bar W) L $  in the linearised approximation. In  the case of the
nonlinear integrand $M \chi^2$ in~\eqref{e:ccF}, this has the result that this integral is
effectively equal to the superspace  integral of $(W \bar W) (  M \chi^2 + \bar M \bar \chi^2 )$ in the
linearised approximation, which is precisely the operator defining the
(parity-even) five-point invariant discussed in~\cite{Beisert:2010jx}.

\section{Conclusion}
\label{sec:conclusion}

In this paper we have seen, perhaps surprisingly, that the volume of  four-dimensional
$\cN$-extended superspace vanishes on-shell. This means that  the leading fully supersymmetric and duality invariant candidate counterterms for the first
ultraviolet  divergences of $\cN\ge4$ supergravity cannot after all be  written  as full  superspace integrals.  

On the other hand, in section~\ref{sec:counterterms} we have exhibited a fully supersymmetric and duality invariant expression for  the ($\csN-1$)-loop $\cN$-extended supergravity counterterm of structure $\nabla^{2(\cssN-4)}R^4$ in the form    of    an   integral over the  $(\csN,1,1)$  harmonic    superspace measure. This measure exists~\cite{Hartwell:1994rp} at the non-linear level  as opposed to the cases of harmonic measures  $(\csN,p,q)$   with either $p>1$ or $q>1$ (for $\cN \ge 5$).  These   invariants  cannot  be   rewritten  as  full superspace integrals at the nonlinear level. For the $\cN=8$ case, the purely gravitational component of this invariant is of the general form
\be 
  I^8\sim \kappa^{12}\,\int d^4x\, e\, \bigl( (\nabla^4 R^2)^2 +{\rm s. s. c. } \bigr) \ .
\ee 
It was  shown in~\cite{Bossard:2011ij} that the absence of a superdiffeomorphism anomaly implies that there exists a duality-invariant form for the associated corrected action $S = S_{\scriptscriptstyle \rm class} + I^8 + \dots$ in the Henneaux--Teitelboim formalism \cite{Henneaux:1988gg}, which is equivalent to the existence of an action satisfying the Gaillard--Zumino constraint in the standard formulation. Duality invariance therefore poses no obstacle to the occurrence of a 7-loop logarithmic divergence, as opposed to what was claimed in \cite{Kallosh:2011dp}.  

There is no known requirement that the  counterterm to an ultraviolet
divergence be given by a full superspace integral with respect to the full on-shell supersymmetry. The situation is similar for counterterms to
the  ultraviolet  divergences   of  maximal  supergravity  in  higher
dimensions, where BPS counterterms, written as subsurface integrals with respect to the full on-shell superspace (at least at the linearised level \cite{Bossard:2009sy}), are known to occur in many cases. For example, the  one-loop  counterterm  in eight  dimensions  is  the $R^4$ invariant  expressed as an on-shell half-superspace  integral, the two-loop $\nabla^4R^4$  
 counterterm in seven dimensions is an on-shell
 quarter-superspace  integral, and the  three-loop $\nabla^6R^4$ counterterm  in six dimensions is an on-shell eighth-superspace
 integral. Off-shell supersymmetry or algebraic renormalisation methods or superstring limiting methods \cite{Howe:2002ui,Bossard:2009sy,Elvang:2010kc,Green:2010sp,Bossard:2010bd,Beisert:2010jx} can rule out certain BPS structures with respect to the full supersymmetry, but none of these methods are known to apply to the $D=4$ seven-loop counterterm \eqref{e:In} for the $\cN=8$ theory, or to the same structure at corresponding loop orders for lesser $\cN$-extended supergravities.

Nonetheless,   the   fact  that   the   invariants  \eqref{e:In}   and
\eqref{e:In2} are not associated to full-superspace integrals might give one pause about their ultimate acceptability as counterterms. One can conceive of further non-renormalisation restrictions that might follow from nonstandard methods. And a full nonlinear analysis of their cocycle structure in the ectoplasm formalism has not yet been carried out. 

There   are   possible   analogues  of   further   non-renormalisation
restrictions in super  Yang--Mills theories. An example concerns
the   absence   of   the   three-loop   double-trace   divergence   in
six-dimensional  $\cN=2$ super Yang--Mills  theory \cite{Bern:2010tq}.
In that  case, the  double-trace invariant $(\partial\,  \mbox{tr} F^2
)^2$ descends from a 1/4  BPS primary operator.  The cocycle structure
of  this invariant  is moreover  identical  to that  of the  classical
action, so  that one  does not at  present have  a non-renormalisation
theorem for it within the framework of algebraic renormalisation. Since the 7-loop maximal supergravity divergence candidate turns out to be the  superspace integral of a G-analytic  superfield, it might have similar properties.  Arguments using the pure spinor formalism in string theory and field theory
\cite{Berkovits:2009aw,Bjornsson:2010wm,Bjornsson:2010wu} show the
super Yang--Mills invariant to be protected beyond the two-loop
order, but these arguments do not, however, carry over straightforwardly to the gravitational case.

In  spacetime dimensions  $D>4$, it seems most  likely that  the full
on-shell  superspace volumes  do  not vanish.  The volume of superspace is only pertinent for higher dimensional logarithmic divergences in the case of $\cN=1$ (half maximal) supergravity in 8 dimensions at one loop, and for $\cN=2$ (maximal) supergravity in nine dimensions at two loops. For  example, the  two-loop, four-graviton amplitude for maximal
$D=9$ supergravity is ultraviolet divergent with  a $\nabla^8 R^4$
counterterm~\cite{Bern:1998ug}.  The  duality-invariant supersymmetric
counterterm  of this  dimension  will be  either  the full  superspace
volume for $D=9$ maximal supergravity or a partial superspace integral
along the lines of Section \ref{sec:counterterms} of this paper. If it
turns out to be the superspace  volume, this would not be in
contradiction with the vanishing  of the $D=4$ superspace volumes that
we have found, however. If a  superspace volume is non-vanishing in a dimension
$D>4$, its  reduction to $D=4$  would lead to  a non-duality-invariant
$D=4$  full-superspace  integral of  some  function  of the  dilatonic
scalars arising from the dimensional  reduction, and not to one of the
duality-invariant  counterterms that  we have  constructed  in Section
\ref{sec:counterterms}.  

For maximal supergravity in $D=5$ the volume is not a possible counterterm. The first possible counterterms that are duality invariant and fully supersymmetric occur at the 6-loop order and are schematically of the form $\nabla^{12} R^4$. These can be expressed as full superspace integrals of dimension 4 superspace integrands constructed from the superspace tensors but with no explicit factors of the scalars.

The duality-invariance  properties of a counterterm  can be classified
by the Laplace equation satisfied by the scalar-field prefactor of the
purely           gravitational          part           of          the
invariant~\cite{Green:2005ba,Bossard:2010bd}.      In     perturbative
supergravity  field  theory,   where  one  requires  invariance  under
continuous  duality   transformations,  the  scalar   prefactor  of  a
duality-invariant
counterterm~\cite{Green:2010wi,Green:2010kv,Elvang:2010kc,Bossard:2010bd,Beisert:2010jx}
as   constructed   in   Section~\ref{sec:counterterms}  must   be   an
eigenfunction  of  the duality-invariant  Laplace  operator with  zero
eigenvalue.

In contrast, at the nonperturbative string-theory level, maximally supersymmetric string-theory considerations indicate~\cite{Green:2008bf} that the scalar prefactors of effective-action contributions such as the dimension-16 $\nabla^8 R^4$ operator will be sums of automorphic forms under the corresponding discrete duality group, arising from solutions to the corresponding Laplace equation with various  eigenvalues. In the field-theory limit, such contributions nonetheless reduce to continuously duality-invariant expressions. For example, it was shown in~\cite{Green:2008bf} that the 2-loop $D=9$ maximal supergravity divergence is contained in the zero-eigenvalue $SL(2,\IZ)$ invariant automorphic contribution to the $\nabla^8 R^4$ operator.

Of course, should duality symmetries be broken by anomalies, they cannot be used to constrain ultraviolet counterterms. This caveat applies in particular to the case of $\cN=4$ supergravity, where quantum  corrections break the corresponding global $SU(1,1)$ symmetry, so that one can  consider a full-superspace integral of any function $F(W \bar W)$ of that theory's complex scalar field $W$ parametrising $U(1)\backslash SU(1,1)$\,,
\be 
I^4_F = \kappa^{4}\, \int d^4x\, d^{16}\theta\, E(x,\theta)\, F(W\bar W)\ ;
\ee 
such  integrals  are  in  general  non-vanishing and will contribute in the
linearised approximation to couplings of the form $F^\ord{2}(\phi\bar \phi)\,R^4$ plus
supersymmetric completions. So one should keep in mind that the strong limitations on the forms of ultraviolet counterterms that we have considered in this paper follow both from supersymmetry and from the requirement of continuous duality invariance where applicable.

\acknowledgments

We  would like  to thank  Niklas Beisert, Nathan Berkovits,  Henriette  Elvang, Daniel
Freedman,  Michael Kiermaier,  Emery  Sokatchev and  Boris Zupnik  for
useful discussions  and comments on  this work. P.V. and  K.S.S. would
like  to  thank  the  Kavli  Institute  for  Theoretical  Physics  for
hospitality during the course of this work, and for support in part by
the    National   Science    Foundation   under    Grant    No.\   NSF
PHY05-51164. K.S.S.  would also  like to thank  the TEO  Department of
CBPF  for  hospitality during  the  course of  the  work,  in a  visit
supported by a PCI-BEV grant. The work of K.S.S. was supported in part
by the STFC under rolling grant ST/G000743/1.  The work of G.B. was supported by the ITN programme PITN-GA-2009-237920, the ERC Advanced Grant 226371, the IFCPAR CEFIPRA programme 4104-2 and the ANR programme blanc NT09-573739.
\bigskip

\appendix
\section{On-shell extended Superspace}\label{sec:superspace}

In  this appendix,  we  review the  main  properties of  $\cN$-extended
superspace in four dimensions 
needed  for  the  computation  in   the  main  text.   We  follow  the
conventions and notation of~\cite{Howe:1981gz}.

At the  nonlinear level, the  solutions to the Bianchi  identities are
expressed  in   terms  of  the  spin  1/2    fermions
$\chi_\alpha^{ijk}$  and  $\chi_{\alpha\,  ijklm}$ and  their  complex
conjugates:

\begin{eqnarray}
\nn  R^i{}_j&=& -\frac13\, \bar P^{iklm} \wedge P_{jklm}\\
\nn P^i_\alpha{}_{jklm}&=&2\delta^i_{[j}\,\chi_{\alpha\, klm]}\ ,\quad 
P_{\dot\alpha i\, jklm}=\bar\chi_{\dot\alpha\, ijklm}\\
D_{\alpha}^i\bar\chi_{\dot\beta\,jklmn}&=& 5i\delta^i_{[j}
P_{\alpha\dot\beta}{}_{klmn]}\ ,\quad 
\nn \bar D_{\dot\alpha\,i} \chi_{\beta\,jkl}=2iP_{\beta\dot\alpha\,ijkl}\\
\nn D_{\alpha}^i \chi_{\beta}^{jklmn}&=&
M_{(\alpha\beta)}^{ijklmn}-\frac52\varepsilon_{\alpha\beta}\,
\bar\chi_{\dot\alpha}^{ i[jk}\bar\chi^{\dot\alpha\,lmn]}\\
\label{e:Dchi} D_{\alpha}^i \chi_{\beta\,jkl}&=&3\delta^{i}_{[j} M_{\a\b\,kl]}+
\varepsilon_{\a\b}\,\Big(\frac{2}{ \csN-4}\,
  \bar\chi_{\dot\alpha\, ijkmn}\bar\chi^{\dot\alpha\,lmn}\\
\nn &&-\frac{3}{    (\csN-3)(\csN-4)}         \,         \delta^l_{[i}\bar\chi_{\dot\alpha\,
    jk]mnr}\bar\chi^{\dot\alpha\, mnr}\Big)\ .
\end{eqnarray}
All $i,j,\dots$ indices are $(S)U(\csN)$ indices.

For $\cN=8$,  we have also  $\bar P^{ijkl}= \frac{1}{24} \varepsilon^{ijklmnpq}
P_{mnpq}$.

It was shown in~\cite{Howe:1981gz} that the fermions $\chi_\alpha^{ijk}$ and $\chi_{\alpha\, ijklm}$ 
arise from the fermionic part of the off-diagonal components of the superspace Maurer--Cartan form for the scalar potential ${\cal V}$ parametrising the
coset space $K\backslash G$ given by $U(4)\backslash \bigl( SU(1,1) \times
SU(4)\bigr)\cong U(1)\backslash SU(1,1)$ for $\cN=4$,
$U(5)\backslash SU(5,1)$ for $\cN=5$, $U(6) \backslash SO^*(12)$ for $\cN=6$ and
$(SU(8)/\IZ_2) \backslash E_{7(7)} $ for $\cN=8$. For $\cN=8$
\be 
  \label{e:dVV}
  d{\cal V}\cdot {\cal V}^{-1}= 
  \begin{pmatrix} \frac{2}{3} \delta^{[i}_{[k} \Omega^{j]}{}_{l]} 
  &P_{ijkl}\cr
\bar P^{ijkl}& - \frac{2}{3} \delta_{[i}^{[k} \Omega^{l]}{}_{j]} 
  \end{pmatrix}\ .
\ee 
For further reference, we define the quantities
\be 
  \label{e:JK2}
 J_{\alpha\dot\beta}{}^{ij}_{kl}=\bar\chi_{\dot\beta}^{ijm}\chi_{\alpha\,
klm}\ ,\qquad 
K_{\alpha\dot\beta}{}^{ij}_{kl}=\chi_{\alpha}^{
ijmnp}\bar\chi_{\dot\beta\, klmnp}
\ee 
and 
\begin{eqnarray}
  \label{e:GH}
  H_{\alpha\bd}{}^i_j&=&
  \begin{cases}
 \frac12\,   J_{\alpha\bd}{}^{ik}_{jk}-   \frac{1}{   16}   \delta^i_j\,
 J_{\alpha\bd}{}^{mn}_{mn}& \mbox{for}~\cN=4,8\cr
\frac12\,   J_{\alpha\bd}{}^{ik}_{jk}-   \frac{1}{  16}   \delta^i_j\,
 J_{\alpha\bd}{}^{mn}_{mn}+ \frac16\,   K_{\alpha\bd}{}^{ik}_{jk}-   \frac{1}{ 80}   \delta^i_j\,
 K_{\alpha\bd}{}^{mn}_{mn}& \mbox{for}~\cN=5,6
  \end{cases}\\
 G_{\alpha\bd}&=&
  \begin{cases}
-  \frac{1}{48}  \,
 J_{\alpha\bd}{}^{mn}_{mn}& \mbox{for}~\cN=4,8\cr
-   \frac{1}{48}   \,
 J_{\alpha\bd}{}^{mn}_{mn}+  \frac{7}{240}   \,
 K_{\alpha\bd}{}^{mn}_{mn}& \mbox{for}~\cN=5,6
  \end{cases}
\end{eqnarray}
and
\be 
  \label{e:N}
  N_{\alpha\beta}^{ij}=
  \begin{cases}
    0 & \mbox{for}~\cN=4\cr
   \frac13\, \chi_{(\alpha}^{ijklm}\chi_{\beta)klm}& \mbox{for}~\cN=5,6\cr
   -\frac{1}{72}\,   \varepsilon^{ijklmnpq}\chi_{\alpha\,   klm}\chi_{\beta
     npq}& \mbox{for}~\cN=8 \,.
  \end{cases}
\ee 
%

\subsection{G-analyticity conditions in $\cN=4$ superspace}
\label{sec:N4superspace}

We can check that $J_{\alpha\bd}{}^{1i}_{4j}$ is G-analytic   because $D^k_\alpha\chi^\alpha_{ijk}=0$ in~\cite[eq.~(5.5)]{Howe:1981gz}:
\be
  \label{e:DJ}
  D_{\alpha}^1 J_{\alpha\dot\beta}{}^{1i}_{4j}=0\ ,\qquad
\bar  D_{\dot\alpha\,4} J_{\alpha\dot\beta}{}^{1i}_{4j}=0
\ee
so $B_{\alpha\bd}=J_{\alpha\bd}{}^{1i}_{4i}$ is G-analytic, as well as 
\be
  \label{e:JJ4}
  C_4=\varepsilon^{\alpha\beta}\varepsilon^{\ad\bd} \, J_{\alpha\dot\beta}{}^{1i}_{4j}J_{\beta\ad}{}^{1j}_{4i}\ .
\ee
However, in $SU(4)$ this  expression for $C_4$ turns out to  be   proportional  to $\varepsilon^{\alpha\beta}\varepsilon^{\ad\bd} B_{\alpha\bd}B_{\beta\ad}$\,. Therefore for $\cN=4$, $B_{\alpha\bd}$ and all its powers are G-analytic.

\subsection{G-analyticity conditions in $\cN=5$ superspace}
\label{sec:cn=5-superspace}

For $\cN=5$, we have that

\be 
  \label{e:KN5}
  K_{\alpha\bd}{}^{1i}_{5\,    j}    =-6    \delta^i_5\delta^1_j    \,
  \chi_\alpha^{12345}\bar\chi_{\bd\, 12345}\ .
\ee 
This implies that $K_{\alpha\bd}{}^{1i}_{5\, i}=0$\,.
Acting with the fermionic derivatives leads to

\begin{eqnarray}\label{e:DN5}
\nn  D_{\gamma}^1
  J_{\alpha\bd}{}^{1i}_{5 j}&=&\frac{1}{6}\, \delta^i_5 \delta^1_j\,\varepsilon_{\gamma\alpha}\bar\chi_{\dot\beta}^{
   1 5p}\,  \bar\chi_{\dd\, 1pqr5}
   \bar\chi^{\dd\,1qr}\\
D_\gamma^1
K_{\alpha\bd}{}^{1i}_{5 j}&=&\delta^i_5\delta^1_j\,\left(-\frac{3}{2}\varepsilon_{\gamma\alpha}\,
\bar\chi_{\ad}^{15p}    \bar\chi^{\ad\,    qr1}\bar\chi_{\bd\,  1
  pqr5}-i6\chi_{\alpha}^{12345} P_{\gamma\bd\, 2345}\right)\ ,
\end{eqnarray}
with equivalent expressions for the action of $\bar D_{\cd\,\cN}$.
These equation imply  that $D_\gamma^1 J_{\alpha\bd}{}^{1i}_{5 i}=\bar
D_{\cd\,5}   J_{\alpha\bd}{}^{1i}_{5    i}=0$\,,   so  
$B_{\alpha\bd}=J_{\alpha\bd}{}^{1i}_{5i}$ is G-analytic.

Since  $J_{\alpha\bd}{}^{11}_{5
  5}=K_{\alpha\bd}{}^{11}_{5 5}=0$\,, we find that
$  J_{\alpha\dot\beta}{}^{1i}_{5j}K_{\alpha\dot\beta}{}^{1j}_{5i}  =0$
and $K_{\alpha\dot\beta}{}^{1i}_{5j} K_{\alpha\dot\beta}{}^{1j}_{5i}=0$\,,
so the only term to analyse at quartic order is
\be 
  \label{e:JJ5}
  C_5=\varepsilon^{\alpha\beta}\varepsilon^{\ad\bd}            \,            
  J_{\alpha\dot\beta}{}^{1i}_{5j}
  \, J_{\beta\ad}{}^{1j}_{5i}\ ,
\ee 
but        $C_5\propto    \varepsilon^{\alpha\beta}\varepsilon^{\ad\bd}
B_{\alpha\bd}B_{\beta\ad}$.
Therefore for $\cN=5$, $B_{\alpha\bd}$ and all its powers are G-analytic.

\subsection{G-analyticity conditions in $\cN=6$ superspace}
\label{sec:n6-superspace}

For $\cN=6$, the $J$ and $K$ fermion bilinears are non-vanishing and are independent.

The variation of these bilinears is given by

\begin{eqnarray}\label{e:DN6}
\nn  D_{\gamma}^1
  J_{\alpha\bd}{}^{1i}_{6 j}&=&\varepsilon_{\gamma\alpha}\bar\chi_{\dot\beta}^{
   1 im}\, \left( \bar\chi_{\dd\, 6 jmrs}
   \bar\chi^{\dd\,1rs}+\frac{1}{                              6}
   \delta^1_{j}\bar\chi_{\dd\, 6 mpqr}\bar\chi^{\dd\,pqr}\right)\\
D_\gamma^1
K_{\alpha\bd}{}^{1i}_{6j}&=&\left(-\frac{3}{2}\varepsilon_{\gamma\alpha}\,
\bar\chi_{\ad}^{1[ip}    \bar\chi^{\ad\,  qr]1}\bar\chi_{\bd\,  jpqr6}-5i\delta_j^1\,\chi_{\alpha}^{1ipqr} P_{\gamma\bd\, pqr6}\right)\ ,
\end{eqnarray}
with equivalent equations for the action of $\bar D_{\cd\,6}$\,.

These    equations    and    the   Fierz    identity    $\theta_\alpha
\psi_\beta\psi^\beta=-2 \theta_\beta\psi^\beta\,\psi_\alpha$ 
imply that 
\begin{eqnarray}
B_{\alpha\bd}&=& J_{\alpha\bd}{}^{1i}_{6 i}+\frac{1}{3} K_{\alpha\bd}{}^{1i}_{6 i}\\
C_6&=& \varepsilon^{\alpha\beta}\varepsilon^{\ad\bd} 
\Bigl( J_{\alpha\bd}{}^{1i}_{6 i} J_{\beta\ad}{}^{1j}_{6 \,j} +\frac{4}{3}\, J_{\alpha\bd}{}^{1i}_{6\,j} J_{\beta\ad}{}^{1j}_{6 \,i} \Bigr) \ .
\end{eqnarray}
Therefore $B_{\alpha\bd}$ and all its powers and $C_6$ are G-analytic.

\subsection{G-analyticity conditions in $\cN=8$ superspace}
\label{sec:N8superspace}

In $\cN=8$, because we have the relations 

\be \label{e:dualchi}
 \bar \chi_\a^{ijklm} = \frac{1}{12} \varepsilon^{ijklmnpq} \chi_{\a\, npq} \; , \qquad \bar \chi_{\ad\, ijklm} =  \frac{1}{12} \varepsilon_{ijklmnpq} \bar  \chi_{\ad}^{npq}  \ ,
\ee 
We find that the G-analyticity conditions lead to 
\begin{eqnarray}\label{e:JGanalyticN8}
\nn  D_{\gamma}^1               J_{\alpha\dot\alpha}{}^{1i}_{8j}&=&- \frac{1}{48}\,
  \varepsilon_{\gamma\alpha}\varepsilon^{\dot\gamma\dot\delta}\,
  \varepsilon_{18ikmnpq}\,                      \bar\chi_{\dot\alpha}^{1jk}
  \bar\chi_{\dot\gamma}^{1mn} \bar\chi_{\dot\delta}^{1pq}\ ,\\
D_{\gamma}^1               J_{\alpha\dot\alpha}{}^{1i}_{8i}&=&0\ , 
\end{eqnarray}
and similarly for the complex conjugate. Therefore $B_{\alpha\bd}$ and all its powers are G-analytic.

\section{Kinematic structure}\label{sec:Kinematics}

Supersymmetry Ward identities imply that the four-graviton amplitude kinematic structure is always of the form 
\be 
P(s,tu)\,C^{\ord{1}}_{\alpha\beta\gamma\delta}                   C^{\ord{2}\,
  \alpha\beta\gamma\delta} \bar C^{\ord{3}}_{\ad\bd\cd\dd} \bar C^{\ord{4}\,\ad\bd\cd\dd} + {\rm c. c.} 
+\mbox{perms}\,(2,3,4)
\ee 
where
\be C^{\ord{n}}_{\alpha\beta\gamma\delta} =  \sigma^{ab}_{(\alpha\beta}   \sigma^{cd}_{\gamma\delta)} \,  k^{\ord{n}}_{a}  k^{\ord{n}}_{c}  \epsilon^{\ord{n}}_{bd} (k^{\ord{n}}) \ee
 is the Weyl tensor associated to the $n^{\rm th}$ graviton of momentum $k^{\ord{n}}$ and polarisation $\epsilon^{\ord{n}}$, and  $\mbox{perm}\,(2,3,4)$ denotes  the  sum over  the  permutations of  the
labels  of  the   external  particles  while  $s=(k^{\ord{1}}+k^{\ord{2}})^2$,
$t=(k^{\ord{1}}+k^{\ord{4}})^2$  and   $u=(k^{\ord{1}}+k^{\ord{3}})^2$  since  we  are working with the signature $(+---)$.  For   the  contribution   of  order
$\nabla^{2k}R^4$, $P_k(s,tu)$ is a polynomial of degree $k$ in $s,t,u$:
\be 
P_k(s,tu)= \sum_{i=0}^{\lfloor k/2\rfloor}  \, c_k^i \, s^{k-2i}\, (tu)^{i}\ .
\ee 
One sees immediately that there are $\lfloor k/2\rfloor+1$ independent monomials at each order.

In the case of $\cN=8$ supergravity, $C_{\alpha\beta\gamma\delta}$ and $\bar C_{\ad\bd\cd\dd}$  occur in the same linearised supersymmetry multiplet, and the supersymmetry Ward identities therefore imply that the dependence   on  the   polarisations  factorises   the  four-graviton
 amplitude such that $P_k(s,tu)$ is a symmetric function in $s,t,u$. $P(s,tu)$ is then expressed as a polynomial in
 the  invariants $\sigma_2=s^2+t^2+u^2 = 2 ( s^2 - tu) $ and  $\sigma_3=s^3+t^3+u^3= 3 stu$ as
 shown in~\cite{Green:1999pv}.  The kinematic structure
 $\nabla^{2k}R^4$ has degeneracy $\lfloor (k+2)/2\rfloor-\lfloor (k+2)/3\rfloor$, and is
 unique for $k=0$, $2\leq k\leq5$ and $k=7$.


\end{document}